\documentclass{aa}
\usepackage{txfonts}
\usepackage{natbib}
\bibpunct{(}{)}{;}{a}{}{,}
\usepackage{graphicx}

\begin{document}
\title{Polychromatic View of Intergalactic Star Formation in NGC~5291}

\author{M\'ed\'eric Boquien\inst{1,2} \and Pierre-Alain Duc\inst{1} \and Jonathan Braine\inst{3} \and Elias Brinks\inst{4} \and Ute Lisenfeld\inst{5} \and Vassilis Charmandaris\inst{6,7,8}}

\offprints{M\'ed\'eric Boquien, \email{mederic.boquien@cea.fr}}

\institute{AIM -- Unit\'e Mixte de Recherche CEA -- CNRS -- Universit\'e Paris VII -- UMR n$^\circ$ 7158
\and CEA-Saclay, DSM/DAPNIA/Service d'Astrophysique, 91191 Gif-sur-Yvette CEDEX, France
\and Observatoire de Bordeaux, UMR 5804, CNRS/INSU, B.P. 89, F-33270 Floirac, France
\and Centre for Astrophysics Research, University of Hertfordshire, College Lane, Hatfield AL10 9AB, UK
\and Dept. de F\'\i sica Te\'orica y del Cosmos, Universidad de Granada, Granada, Spain\and Department of Physics, University of Crete, GR-71003, Heraklion, Greece
\and IESL/Foundation for Research and Technology - Hellas, GR-71110, Heraklion, Greece
\and Chercheur Associ\'e, Observatoire de Paris,  F-75014, Paris, France}

\date{Received / Accepted}

\abstract
{Star formation (SF) takes place in unusual places such as way out in the intergalactic medium out of material expelled from parent galaxies.}
{Whether SF proceeds in this specific environment in a similar way than in galactic disks is the question we wish to answer. Particularly, we address the reliability of ultraviolet, H$\alpha$ and mid-infrared as tracers of SF in the intergalactic medium.}
{We have carried out a multiwavelength analysis of the interacting system NGC~5291, which is remarkable for its extended HI ring hosting numerous intergalactic HII regions. We combined new ultraviolet (GALEX) observations with archival H$\alpha$, 8 $\mu$m \textit{(Spitzer Space Telescope)} and HI (VLA B-array) images of the system.}
{We have found that the morphology of the star forming regions, as traced by the ultraviolet, H$\alpha$, and 8 $\mu$m emission is similar. The 8.0~$\mu$m infrared emission, normalised to emission from dust at 4.5 $\mu$m, which is known to be dominated by PAH bands, is comparable to the integrated emission of dwarf galaxies of the same metallicity and to the emission of individual HII regions in spirals. The 8.0~$\mu$m in the intergalactic environment is therefore an estimator of the star formation rate which is as reliable in that extreme environment as it is for spirals. There is a clear excess of ultraviolet emission compared to individual HII regions in spirals, i.e. the $\left[8.0\right]/\left[NUV\right]$ and $\left[H\alpha\right]/\left[NUV\right]$ SFR ratios are on average low although there are some large variations from one region to another, which cannot be explained by variations of the metallicity or the dust extinction along the HI structure. Comparing the observed SFR with a model of the evolution of $\left[H\alpha\right]/\left[NUV\right]$ with time favours young, quasi-instantaneous though already fading starbursts. The total star formation rate measured in the intergalactic medium (which accounts for 80\% of the total) surrounding NGC~5291 is up to 1.3 M$_\textrm{\sun}$\,yr$^{-1}$ -- a value typical for spirals -- assuming the standard SFR calibrations are valid. The SFR drops by a factor of 2 to 4 in case the star formation is indeed quasi-instantaneous.
}
{}

\maketitle

\section{Introduction}

It has long been known that ``vast and often very irregular swarms of stars and other matter exist in the spaces between the conventional spiral, elliptical, and irregular galaxies'' \citep{zwicky1951a}. Those stars have been stripped off from their host galaxies during interactions, the intergalactic medium being too diffuse and hot to allow in situ star formation. However a fraction of the stellar population could nevertheless have been formed locally out of the gas which has been pulled out from peripheral regions of gas rich galaxies. This has been shown to occur in simulations of tidal encounters \citep{hibbard1995a,duc2004a} as well as in the case of collisional rings \citep{appleton1996a,horellou2001a}. This gas which has the metallicity of the parent galaxy (roughly solar metallicity in the case of normal star forming spirals\footnote{Taking $12+\log{(O/H)}=8.66$ for the solar metallicity \citep{asplund2005a}.}, lower for dwarf galaxies, see \cite{mendes2006a} for instance) can then be reprocessed to form new stars and even new dwarf-like galaxies if the expelled gas reservoir was massive enough. How star formation (SF) proceeds in collisional debris between galaxies, in an environment which is isolated from the direct influence of the galactic disks, and how it compares with more conventional galactic star formation has so far not been studied in detail.

The number of reported ``intergalactic'' star forming regions has grown rapidly in the last few years, especially in groups \citep{mendes04} and clusters of galaxies \citep{cortese06}. The area around NGC~5291 -- an interacting system located at the edge of the Abell 3574 cluster -- is one of the most spectacular objects where numerous instances of intergalactic HII regions are found.

The system is composed of an early-type galaxy, NGC~5291, and a highly disturbed companion galaxy, ``the Seashell''. The first detailed observation of the system was carried out by \cite{longmore1979a} who detected on each side of NGC~5291 blue clumps identified as giant extragalactic HII regions. The spectra of some of them were obtained thanks to an object prism survey of HII galaxies \citep{maza1991a,pena1991a}. \cite{malphrus1997a} used the Very Large Array to map the neutral gas of the system. They discovered a huge asymmetrical ring-like structure connected to NGC~5291. It turned out to be one of the two most gas rich objects in the Southern hemisphere \citep{koribalski04}. The HI structure consists of a series of sub-condensations which correspond to the previously identified intergalactic HII regions.

\cite{duc1998a} published a detailed optical (imagery and spectra) and near-infrared study of the system. They found that the HII regions had moderately high metallicity ($12+\log{(O/H)}=8.4-8.6$) and could thus exclude a primordial origin for the HI structure. They claimed that the ring structure resulted from the past collision of an extended HI disk with a companion. They also showed that the most recent starbursts along the ring are younger than $5\times10^6$ years without any evidence for an underlying old stellar population. \cite{braine01} obtained with the SEST antenna CO$(1-0)$ and CO$(2-1)$ spectra of two of the brightest condensations. Inferring the molecular gas content from the CO emission and comparing it with the SFR determined from H$\alpha$, they determined that their star formation efficiency (SFE) was closer to that measured on average in spiral galaxies than in dwarfs of the same luminosity. Incidently, the detection of the millimetre lines was a further indication that the gas around NGC~5291 had been previously enriched. Finally, \cite{higdon2006a} performed mid-infrared observations, both imaging and spectroscopy, using the IRAC and IRS instruments on the {\em Spitzer Space Telescope}. They unambiguously detected Polycyclic Aromatic Hydrocarbon (PAH) features towards the two brightest condensations of the system. Their spectral analysis also confirmed the young ages of the star forming episodes.

In order to study in more detail the characteristics of the SF regions around NGC~5291, we collected ultraviolet images from the GALEX satellite, ground-based H$\alpha$ images, and high resolution HI maps obtained with the Very Large Array which we combined with optical and mid-infrared {\em Spitzer} images of the system. See Figure \ref{fig:compose} for a combination of those images.

In section \ref{sec:obs-reduction} we present the multiwavelength data set we have used. In particular, we describe how we defined and selected the individual intergalactic star forming regions and how we performed accurate aperture photometry on them. In section \ref{sec:results}, we describe their main properties: overall distribution and star formation rates inferred from different tracers. We then compare them with those measured in various samples of galaxies and individual star forming regions in spiral disks. In section \ref{sec:discussion}, we discuss the origin of the discrepancies between the various SF indicators and present the possible role of extinction, metallicity and starburst age. We finally speculate on the possible impact of intergalactic star formation on its environment.

Throughout this paper we use $H_0=72$ km\,s$^{-1}$\,Mpc$^{-1}$ which results in a distance for NGC~5291 of 62 Mpc. At this distance, 1\arcsec=0.30 kpc.

\section{Observations and data reduction\label{sec:obs-reduction}}

\subsection{Multi-wavelength data}

We describe in this section the various observations used for this paper. We present for the first time an ultraviolet map obtained with the GALEX space telescope. The high resolution HI maps were taken from the paper by \cite{bournaud2007a}. The ESO/3.6m H$\alpha$ data and {\em Spitzer}/IRAC mid-infrared images were obtained from the literature or from the telescope archives.

\subsubsection{Ultraviolet images}

On May 2005 GALEX was used to acquire a set of NUV (near-ultraviolet; $\lambda_\textrm{eff}=227.1$ nm and $\Delta\lambda=73.2$ nm) broadband images of the target with a total exposure time of 2886 seconds. The field of view has a diameter of 1.24\degr, much larger than the NGC~5291 system. The PSF (point spread function) has a width of $\sim$5.0\arcsec. Due to technical difficulties at the time of observation, no far ultraviolet image (FUV) was obtained.

\subsubsection{H$\alpha$ data}

H$\alpha$ observations were carried out at the 3.6-m telescope of the European Southern Observatory (ESO) on April 2002, using the Fabry-Perot interferometer CIGALE. The seeing was $\sim1$\arcsec. The zeroth moment maps were extracted from the datacubes. For further technical details about the observations and data reduction, see \cite{bournaud2004a}.

Due to the small field of view of the camera, three individual images were acquired, covering respectively the area centred on the most luminous HII region to the south of NGC~5291, the central area close to NGC~5291 and the Seashell galaxies, and the area centred on the most luminous HII region to the north of NGC~5291. The southern and central images overlap partially but there is a gap between the central and northern images, as shown in Figure \ref{fig:compose}.

\subsubsection{Mid infrared images}

The mid-infrared broadband images have been retrieved from the {\em Spitzer} public archives. The observations were carried out on February 2004 with the Infrared Array Camera (IRAC) on the {\em Spitzer Space Telescope} at 3.6 $\mu$m, 4.5 $\mu$m, 5.8 $\mu$m and 8.0~$\mu$m. The resolution is $\sim$3\arcsec. The data have been preprocessed using the {\em Spitzer} Science Center (SSC) pipeline version S11.0.2. Due to the limited field of view (about 5\arcmin), the final image is a combination of two different pointings (each using a Reuleaux dither pattern), separated by 180\arcsec\ (see Figure \ref{fig:compose}). For further details about {\em Spitzer} data processing see the {\em Spitzer} Observing Manual\footnote{http://ssc.spitzer.caltech.edu/documents/som/} and the paper by \cite{higdon2006a}.

\subsubsection{HI observations}
The field containing NGC\,5291 was observed with the NRAO\footnote{The National Radio Astronomy Observatory is a facility of the National Science Foundation operated under cooperative agreement by Associated Universities, Inc.} Very Large Array (VLA) in the 21--cm line of atomic neutral hydrogen. The observations and data reduction are fully described elsewhere \citep{bournaud2007a}. Here we will only dwell on the characteristics of the maps that were used in the analysis in this paper.

We covered the entire velocity range of the system at 48.8\,kHz ($\sim10.45$\,km\,s$^{-1}$), after online Hanning smoothing. The AIPS\footnote{The Astronomical Image Processing System is a software package provided by the National Radio Astronomy Observatory.} package was used to calibrate and Fourier transform the data. For this paper we used natural weighting which results in maps at $9.4 \times 5.6$\,arcsec$^2$ resolution (and a beam position angle of 13\degr). The $1\,\sigma$ rms noise is 0.33\,mJy (corresponding to 3.8\,K).

As the emission in this system is extended over more than 10\,arcmin, some flux due to a smooth, extended HI component might be missing from the interferometer data.  It should be stressed that for spatial frequencies corresponding to scales smaller than $\sim5$\arcmin, all HI is recovered in our maps and that any contribution from an extended component becomes negligible in the small regions studied in this paper.

\subsection{Aperture photometry}

\begin{figure*}
\centering
\includegraphics[width=17cm]{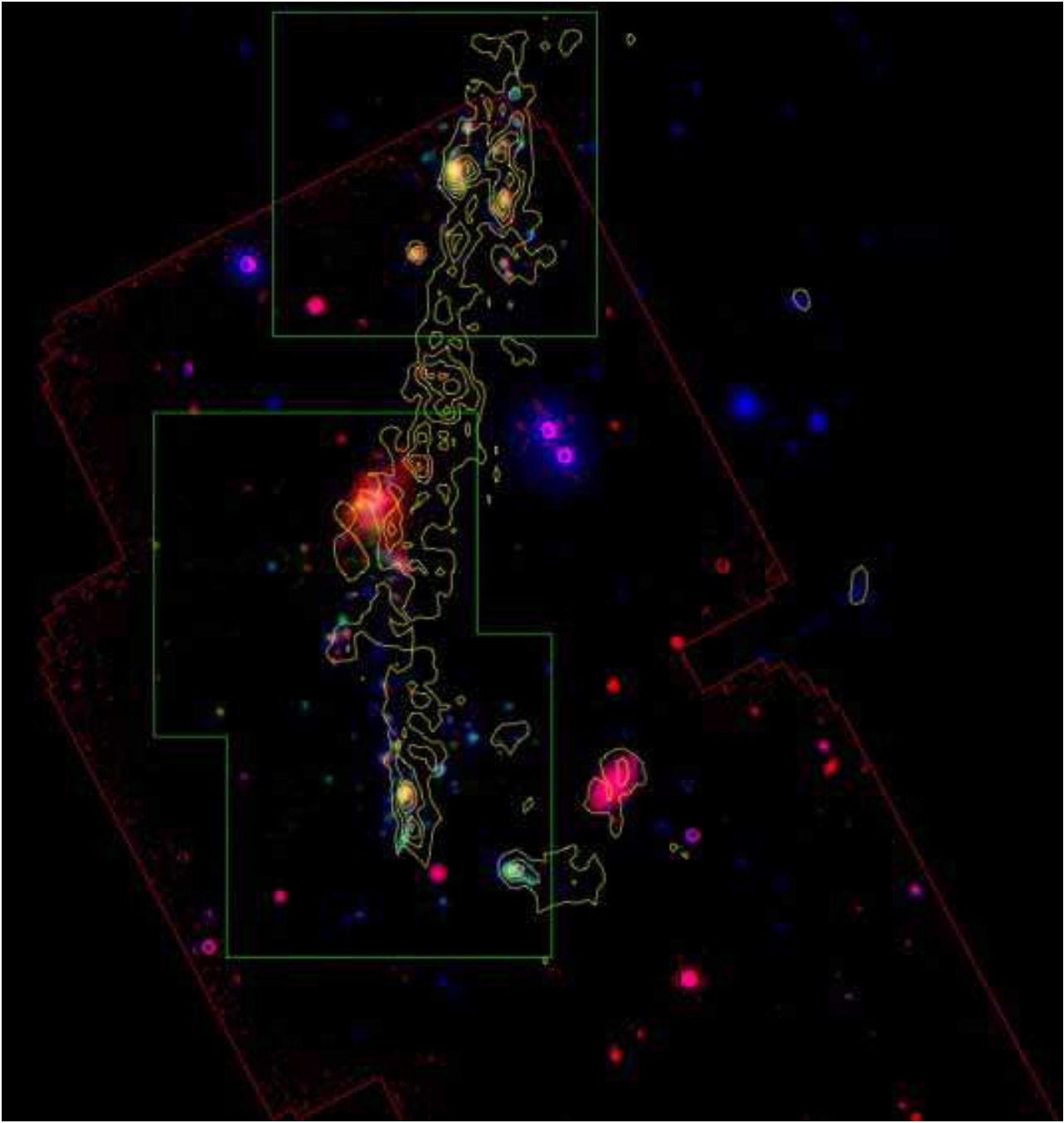}
\caption{Composite image of the NGC~5291 system in pseudo-colours: 8.0~$\mu$m IRAC (red channel), H$\alpha$ (green channel) and near ultraviolet GALEX (blue channel). The overplotted red and green contours represent respectively the 8.0~$\mu$m and the H$\alpha$ fields of view. The blue contours represent the HI emission from the VLA HI map. The levels represent HI column densities of: $5.0\times10^{20}$, $1.1\times10^{21}$, $1.7\times10^{21}$, $2.3\times10^{21}$, $2.9\times10^{21}$, $3.5\times10^{21}$ cm$^{-2}$.\label{fig:compose}}
\end{figure*}

\subsubsection{Selection of the regions\label{sec:selection-regions}}

To identify the intergalactic star-forming regions belonging to the NGC~5291 system and reject from the subsequent analysis unrelated objects we used a combination of optical \citep[B, V and R bands from][]{duc1998a}, H$\alpha$ Fabry-Perot, mid infrared (3.6, 4.5 and 8.0~$\mu$m) and HI observations. We used the following criteria in this order:
\begin{itemize}
\item detection of H$\alpha$ line emission at the NGC~5291 velocity on the first moment map of the Fabry-Perot datacube \citep[see][]{bournaud2004a},
\item proximity to the HI ring-like structure around NGC~5291, as this is the gas reservoir for star formation,
\item presence of diffuse emission in the near-ultraviolet map,
\item presence of extended, diffuse, blue regions on the true-colour BVR optical images, allowing the identification and subsequent removal of bright stars and distant background galaxies,
\item further rejection, based on their colours, of stars and background galaxies using the image obtained combining the three mid-infrared bands.
\end{itemize}

Because the maps obtained in the various wavelength bands do not cover completely the same area, we could in some regions only apply the HI and ultraviolet criteria. The H$\alpha$ map, which of course provides an unambiguous detection of HII regions (apart from spurious lines falling right into the H$\alpha$ redshifted wavelength), unfortunately does not cover all of the HI structure.

As HII regions are irregularly shaped and the system is crowded, we have chosen to define polygonal apertures. We could thus simultaneously take into account the variation of the PSF from one image to another avoiding as much as possible pollution from stars and from nearby objects, which would not have been easily possible using circular apertures. The shape and angular sizes of the polygonal apertures are the same for all bands and thus cover exactly the same physical region.

\begin{figure*}
\centering
\includegraphics[width=17cm]{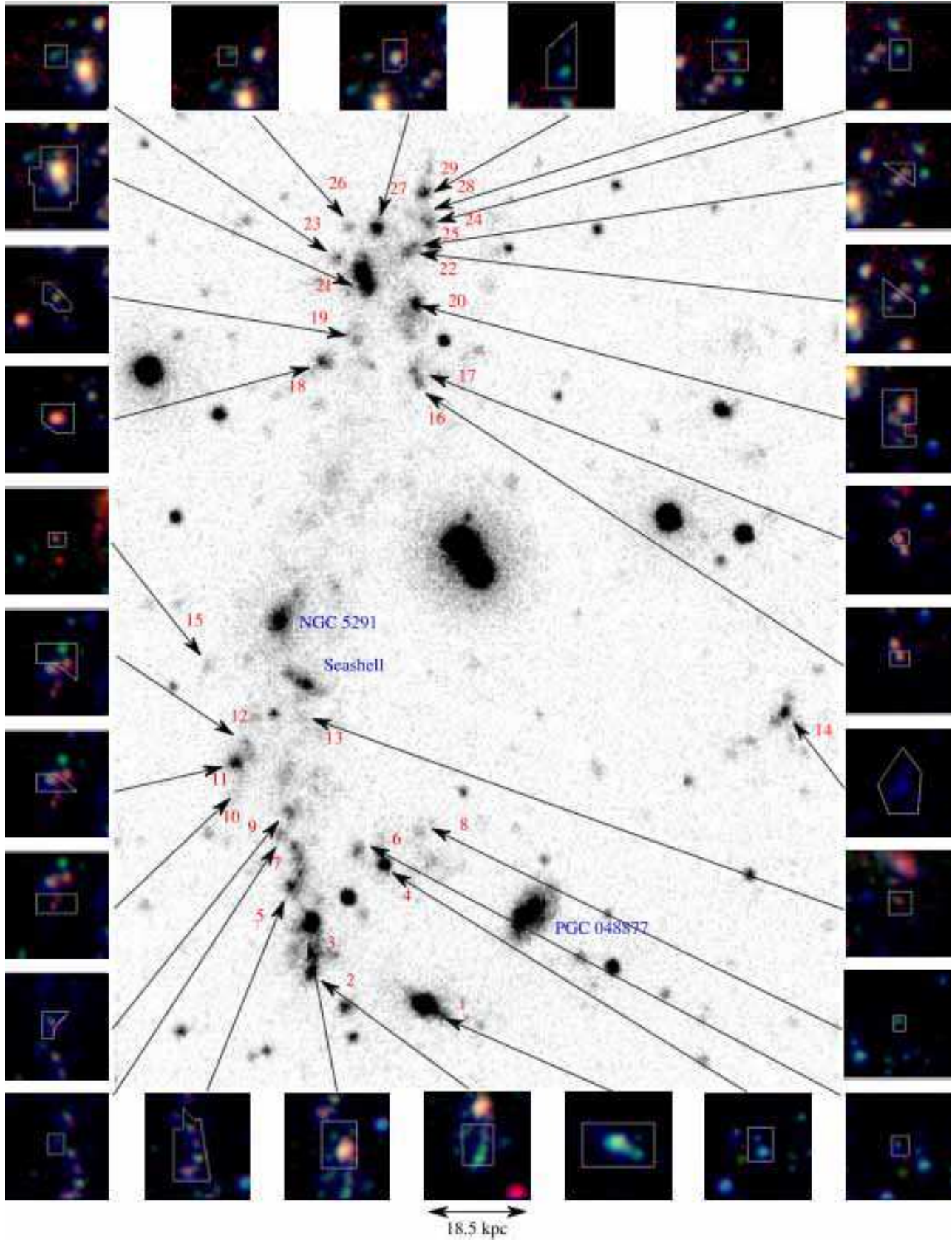}
\caption{Ultraviolet image of NGC~5291 with a mosaic of the selected regions. The individual regions and the main galaxies in the group are labelled. The thumbnails are zoomed images of each selected region using pseudo-colours (red representing 8.0~$\mu$m emission, green H$\alpha$ and blue ultraviolet), with the corresponding polygonal aperture overplotted in white. Note that for region 14 only the UV emission is available and shown (see Fig. 1).\label{fig:vign}}
\end{figure*}

Based on the above mentioned criteria, we have selected 29 regions across the field, which correspond to intergalactic star-forming regions. They are indicated on Figure \ref{fig:vign}.

For reference, we have also measured the fluxes towards the parent galaxies (NGC~5291 and the Seashell) and towards an HI peak just North of NGC~5291 with only faint ultraviolet and mid-infrared counterparts (information on H$\alpha$ is lacking there). 

\begin{figure*}[!htbp]
\centering
\includegraphics[width=17cm]{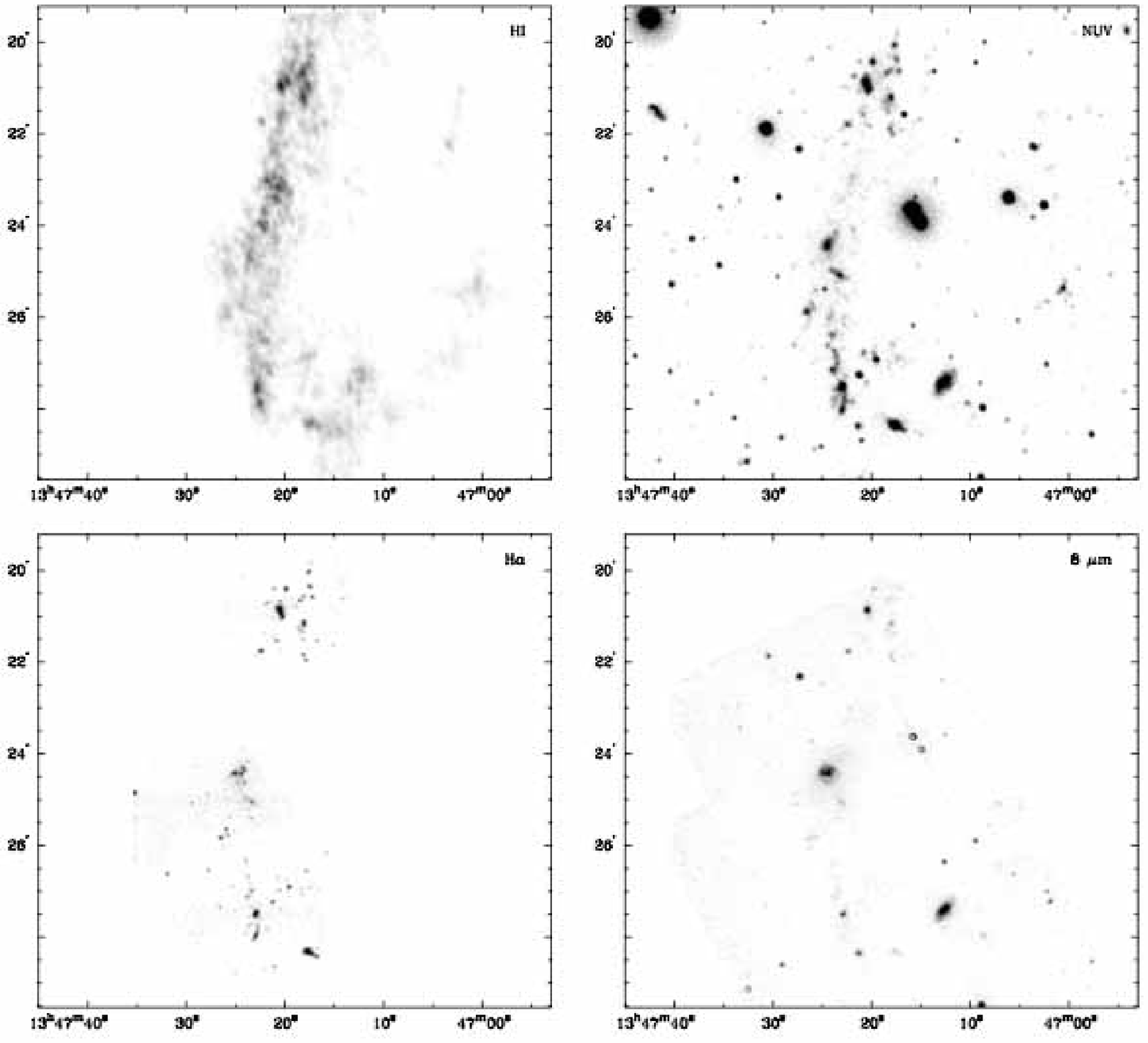}
\caption{The different tracers of star formation towards NGC~5291: VLA HI map (top left), GALEX near-ultraviolet (top right), Fabry-Perot H$\alpha$ (bottom left), and {\em Spitzer} 8.0~$\mu$m (bottom right) maps, all at the same spatial scale. The scaled 3.6 $\mu$m emission was subtracted from the 8.0~$\mu$m map to remove contaminating stellar light from the mid-infrared emission \citep[see for instance][]{pahre2004a} and thus mainly shows dust emission.\label{fig:maps}}
\end{figure*}

\subsubsection{Measurements\label{sec:measurements}}

First of all, we have substracted the background in each band. To do so, we have manually measured the background level on each image using the IRAF\footnote{IRAF is distributed by the National Optical Astronomy Observatories, which are operated by the Association of Universities for Research in Astronomy, Inc., under cooperative agreement with the National Science Foundation.} (Image Reduction and Analysis Facility) {\sc imexa} procedure. Depending on the image, a single background was determined in case the background level was flat throughout the image (the ultraviolet image falls in this category as the standard pipeline already performs a background subtraction). Individually determined background levels had to be used for each of the regions in the H$\alpha$ and infrared images. Each level is evaluated as the mean of several (up to 30) measurements of the background at different locations around the selected regions. The uncertainty on the background is the standard deviation.

We have also tested background removal using SExtractor \citep{bertin1996a}. However, the large scale diffuse luminosity made the determination of the level quite uncertain and a small variation of the background extraction parameters caused a large variation of the flux, especially for the faintest regions where the difference between two extractions could be up to a factor of 3.

The standard IRAC mid-infrared pipeline calibrates the images in such a way that the flux of a point source is correctly determined when using a 10 pixel radius circular aperture and calculating the background in an annulus with inner and outer radii of 10 and 20 pixels\footnote{See http://ssc.spitzer.caltech.edu/irac/dh/}. This is due to the fact that the PSF is quite large, especially in the 8.0~$\mu$m band where a significant part of the flux is scattered on arcminute size scales, which is much larger than 10 pixels. The calibrated units are set in such a way that the off aperture flux loss is exactly compensated when it is determined with a 10 pixel circular aperture. Therefore, when the aperture is larger (smaller) than 10 pixels of radius, the flux density is overestimated (underestimated). However in this paper we do not use a circular aperture for the determination of the fluxes. Instead, the areas of interest are enclosed using polygons. As a result another method needs to be applied in order to estimate the flux outside our apertures and perform the necessary corrections. To do so we use the standard correction values published in the IRAC manual for a point source correction, taking an equivalent radius of R=$\sqrt{\textrm{S}/\pi}$, S being the surface in pixel units. The equivalent background annulus was taken with a radius from 10 to 20 pixels. Corrections for aperture areas falling between the published values have been interpolated. The uncertainty on the aperture correction therefore depends on the aperture and the HII region shapes. The Poisson photon noise has also been taken into account. The absolute photometric calibration, according to the IRAC data handbook, is better than 5\%.

As the Fabry-Perot observations were not flux calibrated, we made an indirect calibration of the H$\alpha$ data using the optical multi-object spectroscopic observations taken on July 1994 on the NTT ESO telescope at La Silla \citep{duc1998a}. The calibration is accurate to only about 30\% due to uncertainties on the slit positions. The overlap of the two southern fields of view has been used to check the consistency of the calibration between the individual Fabry-Perot pointings.

The absolute photometric uncertainty of the GALEX near ultraviolet image \footnote{See \cite{morrissey2005a} and the GALEX Observers Guide\newline http://www.galex.caltech.edu/DATA/gr1\_docs/\newline GR1\_Observers\_guide\_v1.htm for more details.} is 7\%.

The fluxes derived in each band -- when available -- are presented in Table \ref{tab:fluxes}. The mid-infrared fluxes obtained have been compared whenever possible with the ones published by \cite{higdon2006a}. The integrated fluxes of NGC~5291, the Seashell and PGC~048877 are consistent to within a few percent. Our photometry of individual star-forming regions agrees with the earlier values to within $\sim$30\% for the case of bright isolated regions, but it deviates substantially for the fainter ones. Our analysis indicates that the differences mainly originate from the way apertures were selected and how the background substraction was performed: automatically by SExtractor in the Higdon et al. study, manually in this paper to make sure that regions with the same physical origin were enclosed.

\section{Results\label{sec:results}}

\subsection{Distribution of the star forming regions}
The individual HI, NUV, H$\alpha$ and 8.0~$\mu$m maps are presented in Figure \ref{fig:maps} and a combination of them in Figure \ref{fig:compose}. They show a remarkably similar morphology, indicating that all of them are physically linked and associated with the same process, most probably star-formation. We observe that star-forming knots are spread over $\sim$180 kpc, all along the HI structure. They are actually located on or close to individual HI peaks.

The brightest regions (in H$\alpha$, NUV and 8.0~$\mu$m) are located near the apparent tip and points of maximum column density of the HI ring: they correspond to our regions 1, 3 and 21; the latter two, also known as NGC~5291S and NGC~5291N, have been discussed in the literature as the possible progenitors of dwarf galaxies, and are probably associated with gravitationally bound gas clouds \citep{bournaud2004a}. As shown in Figure \ref{fig:maps}, the HI condensation just North of NGC~5291, which actually has one of the highest HI column density in the entire system ($N(HI)\ge3.3 \times 10^{21}$ cm$^{-2}$, hereafter referred to as HI concentration) shows only faint diffuse ultraviolet and mid-infrared counterparts\footnote{This region falls outside the H$\alpha$ field of view. In the optical the region is also faint, the surface brightness averaged over the condensation being $\sim$25.2 mag/\arcsec$^2$ in the B band. The global morphology of this region is similar to what is observed in the ultraviolet.}. The UV/HI and IR/HI flux ratios measured within this condensation deviate by an order of magnitude from the values determined from the other regions. One possibility would be that the starburst is extremely young and whatever O and B stars formed there have not dispersed their surrounding dense cloud material yet. Having a CO spectrum of that region and a far-infrared/millimetre image would help in checking this hypothesis. Having said that, inspection of individual HI spectra across this cloud indicates that HI is found spread over a much wider range in velocities than at the other locations investigated. This could mean that what we are seeing in Figure \ref{fig:maps} is a blend of clouds along the line of sight, adding up to result in an apparent massive HI concentration whereas each contributing cloud is much less massive, one of which being just massive enough to have started to form stars.

Figure \ref{fig:vign} presents combined NUV, H$\alpha$ and 8.0~$\mu$m emission of the individual star-forming knots. The false-colour images, all displayed with the same intensity scale, indicate that although they have the same morphology in each band, they show large variations in their relative fluxes. Local variations of the properties of the star forming regions are thus important, despite the fact that they all belong to a single structure probably formed simultaneously during a collision.

Another region, labelled number 1 in Figure \ref{fig:vign}, also shows a level of activity much higher than what would be expected from its HI column density. However the HI distribution of this object shows clear distortions: a gaseous tail without any 8.0~$\mu$m counterpart is seen just to the West of the star-forming region (see Figure \ref{fig:compose}). There is evidence that in addition to the gravitational effects of the collision, the objects in the system may have suffered from the interaction of their interstellar medium with the intergalactic medium belonging to the cluster through which they move. A systematic offset between UV/H$\alpha$/MIR and HI peaks is observed with the HI being shifted to the opposite side of the cluster centre, which is located to the East of the galaxy. Why this process would be more efficient for object 1, which belongs to the same global structure as the other objects is however unclear.

\begin{table*}

\caption{Fluxes of selected regions in the ultraviolet, H$\alpha$, and mid-infrared (3.6, 4.5, and 8.0~$\mu$m). The ``--'' symbol indicates that the region falls outside the field of view, ``$\le$'' means the aperture is heavily polluted by stars and the flux is therefore likely being overestimated and ``$\ge$'' means the aperture does not contain all of the object, e.g., to avoid a foreground object, the actual flux then being underestimated.}
\label{tab:fluxes}
\centering
\begin{tabular}{c c c c c c c c}
\hline\hline
Region&GALEX name&DM98&F$_\textrm{NUV}$&F$_{\textrm{H}\alpha}$&F$_\textrm{IR3.6}$&F$_\textrm{IR4.5}$&F$_\textrm{IR8.0}$\\
&(IAU name)&&($\mu$Jy)&($10^{-18}$W\,m$^{-2}$)&($\mu$Jy)&($\mu$Jy)&($\mu$Jy)\\\hline
1&J134717.57-302821.11&&274$\pm$20&33.2$\pm$10.0&331$\pm$101&254$\pm$35&640$\pm$99\\
2&J134722.91-302750.19&&92$\pm$7&10.7$\pm$3.3&144$\pm$41&98$\pm$23&629$\pm$41\\
3&J134723.00-302724.31&a&150$\pm$11&18.8$\pm$5.7&385$\pm$24&251$\pm$32&2088$\pm$55\\
4&J134719.76-302648.77&&53$\pm$4&4.3$\pm$1.3&131$\pm$12&118$\pm$15.4&207$\pm$27\\
5&J134723.86-302646.25&c&98$\pm$7&5.9$\pm$1.9&221$\pm$28&179$\pm$35&805$\pm$63\\
6&J134720.69-302642.66&&14$\pm$2&1.5$\pm$0.5&31$\pm$5.4&28$\pm$7&97$\pm$12\\
7&J134724.51-302635.91&&14$\pm$1&0.6$\pm$0.2&53$\pm$12&44$\pm$5&73$\pm$12\\
8&J134717.79-302634.82&&6$\pm$1&1.0$\pm$0.3&31$\pm$7&31$\pm$3&60$\pm$7\\
9&J134723.95-302622.21&d&23$\pm$2&1.5$\pm$0.5&84$\pm$15&61$\pm$6&176$\pm$15\\
10&J134726.10-302602.63&&14$\pm$1&1.4$\pm$0.5&112$\pm$26&95$\pm$10&278$\pm$26\\
11&J134726.66-302552.06&e&34$\pm$3&2.5$\pm$0.8&199$\pm$18&138$\pm$8&384$\pm$18\\
12&J134726.45-302545.78&&31$\pm$3&4.3$\pm$1.3&229$\pm$34&169$\pm$13&509$\pm$34\\
13&J134723.37-302524.51&&12$\pm$1&1.3$\pm$0.4&188$\pm$18&135$\pm$8&311$\pm$19\\
14&J134700.56-302518.62&&62$\pm$5&--&--&--&--\\
15&J134727.77-302504.89&&5$\pm$1&0.8$\pm$0.3&70$\pm$9&50$\pm$5&189$\pm$10\\
16&J134717.92-302158.54&f&14$\pm$1&1.4$\pm$0.4&50$\pm$10&41$\pm$6&286$\pm$12\\
17&J134718.08-302150.30&&12$\pm$1&1.3$\pm$0.4&52$\pm$9&37$\pm$6&377$\pm$11\\
18&J134722.47-302144.13&g&30$\pm$2&3.3$\pm$1.0&146$\pm$28&104$\pm$13&1047$\pm$30\\
19&J134720.94-302131.97&&18$\pm$1&1.6$\pm$0.5&96$\pm$18&51$\pm$7&267$\pm$18\\
20&J134717.91-302121.48&h&87$\pm$6&7.8$\pm$2.4&287$\pm$58&244$\pm$22&1428$\pm$58\\
21&J134720.86-302055.70&i&259$\pm$18&46.3$\pm$13.9&638$\pm$38&534$\pm$51&5236$\pm$90\\
22&J134718.18-302042.76&&30$\pm$2&2.1$\pm$0.7&63$\pm$27&34$\pm$16&427$\pm$27\\
23&J134721.66-302041.66&&17$\pm$1&1.5$\pm$0.5&14$\pm$17&no detection&114$\pm$17\\
24&J134717.07-302035.59&&21$\pm$2&1.9$\pm$0.6&1.7$\pm$20&25$\pm$11&31$\pm$20\\
25&J134717.97-302031.99&&15$\pm$1&1.2$\pm$0.4&38$\pm$13&29$\pm$8&235$\pm$14\\
26&J134721.14-302026.07&&9$\pm$1&0.7$\pm$0.2&--&--&--\\
27&J134719.74-302025.73&j&44$\pm$3&4.2$\pm$1.3&116$\pm$23&100$\pm$14&614$\pm$24\\
28&J134717.69-302018.20&&33$\pm$3&2.3$\pm$0.8&--&--&--\\
29&J134717.81-301958.85&k&41$\pm$3&2.5$\pm$0.9&--&--&--\\\hline
Seashell&J134723.17-302504.23&&98$\pm$7&5.6$\pm$1.9&14412$\pm$84&8734$\pm$40&5319$\pm$78\\
NGC~5291&J134717.81-302424.87&&$\ge$293$\pm$24&$\ge$26.2$\pm$9.0&$\ge$48619$\pm$385&$\ge$29195$\pm$141&$\ge$24846$\pm$374\\
HI concentration&J134721.02-302311.33&&71$\pm$8&--&$\le$704$\pm$177&$\le$618$\pm$60&1520$\pm$179\\
\hline
\end{tabular}
\end{table*}

\subsection{Star formation rates in the intergalactic medium\label{sec:SFRIGM}}

The ultraviolet, H$\alpha$ and infrared 8.0~$\mu$m emission are all known to be tracers of the level of star forming activity. The reliability of measuring the star formation rate based on each of these indicators, or a combination of them, has been extensively debated in the literature. It seems to depend very much on the type of environments as many external factors may affect it: contamination by old stars, level of dust extinction, etc. These tracers have not yet been calibrated for the special environments of collisional debris and IGM. We will thus, as a first step, use the classical calibrations obtained for spiral disks to get estimates of the SFR. From a comparison between the various estimates, we will then both discuss the validity of such tracers and see whether some global properties, such as the starburst age may be constrained.

UV radiation is mainly emitted by massive stars younger than $100\times10^6$ years. The star formation rate has been calculated using the standard \cite{kennicutt1998a} relation: SFR(UV)=$\left[NUV\right]=1.4\times10^{-21}$L$_\nu$ [W\,Hz$^{-1}$]\,M$_\textrm{\sun}$\,yr$^{-1}$. This estimator has been built under the assumptions of continuous star formation, with a Salpeter initial mass function \citep[IMF;][]{salpeter1955a} and mass cut-offs from 0.1 to 100 M$_\textrm{\sun}$. It is sensitive to extinction which reprocesses a fraction of the UV emitted by young stars which is reemitted in the infrared.

Ionizing radiation ($\lambda<91.2$ nm) is mainly emitted by very massive, short-lived young stars. This radiation ionizes surrounding gas clouds, which reemit partly in H$\alpha$, thus tracing starbursts. We have used again \cite{kennicutt1998a}: SFR(H$\alpha$)=$\left[H\alpha\right]=7.9\times10^{-35}$L(H$\alpha$) [W]\,M$_\textrm{\sun}$\,yr$^{-1}$, with the same IMF as used previously. This estimator is sensitive to absorption of the ionizing photons by the molecular clouds surrounding young stars and thus very dependent on geometry\footnote{The radiation shortward of 91.2 nm is more affected by dust extinction than the near ultraviolet, therefore H$\alpha$ is more sensitive than ultraviolet to the effects of geometry.}.

At 8.0~$\mu$m we observe a combination of PAH broad emission bands, continuum emission from hot VSG (very small grains) and stellar continuum. The photospheric stellar contribution at 8.0~$\mu$m is lower than 10\% in our special environment which is mostly devoid of an old stellar component (see below). The PAH are heated stochastically by far UV radiation from massive young stars and as such trace star formation. \cite{wu2005a} built a SFR estimator based on a correlation between the 8.0~$\mu$m dust emission (stellar contribution being substracted using the 3.6 $\mu$m band) and radio or H$\alpha$ luminosities. The star formation rate calibration derived by them using the 8.0~$\mu$m/radio correlation is: SFR(IR)=[8.0]=$1.88 \times 10^{-36}\,\nu$L$_\nu\left(8.0\mu\textrm{m}\right)[W]$\,M$_\textrm{\sun}$\,yr$^{-1}$. The Chary-Elbaz templates \citep{chary2001a} and the \cite{kennicutt1998a} estimator based on the total infrared luminosity produce consistent results.

The star formation rates have been calculated using the relations given above and the fluxes presented in Table \ref{tab:fluxes}. We have corrected the fluxes for galactic extinction both in UV and in H$\alpha$ using the Cardelli extinction curve \citep{cardelli1989a}, and taking E(B-V)=0.064 from NED (NASA/IPAC Extragalactic Database). The systematic uncertainty of each estimator has not been taken into account. The UV SFR, $\left[H\alpha\right]/\left[NUV\right]$ and $\left[8.0\right]/\left[NUV\right]$ for each region is presented in Table \ref{tab:sfr}.

We stress that those estimators are likely not very adapted for our case \citep[we expect a very young starburst, see][]{higdon2006a}. However they give more intuitive values to study the variations of the ratio of the SFRs in two bands, from one region to another. We use SFR(UV) as the reference since it is the most direct estimator (unprocessed star light) and more importantly, this is the only estimator which samples the star formation emission along the whole HI structure.

\begin{table*}
\caption{Star formation rates deduced from the ultraviolet, H$\alpha$, and mid infrared, in solar mass per year. The symbol ``:'' means the estimation is uncertain.}
\label{tab:sfr}
\centering
\begin{tabular}{c c c c}
\hline\hline
Region&SFR(UV)&$\left[H\alpha\right]/\left[NUV\right]$&$\left[8.0\right]/\left[NUV\right]$\\\hline
 1&0.247$\pm$0.018&0.57$\pm$0.15&0.09$\pm$0.01\\
 2&0.083$\pm$0.006&0.55$\pm$0.15&0.25$\pm$0.02\\
 3&0.144$\pm$0.010&0.56$\pm$0.15&0.47$\pm$0.03\\
 4&0.047$\pm$0.003&0.39$\pm$0.10&0.14$\pm$0.02\\
 5&0.088$\pm$0.006&0.29$\pm$0.08&0.30$\pm$0.03\\
 6&0.012$\pm$0.001&0.53$\pm$0.17&0.26$\pm$0.04\\
 7&0.012$\pm$0.001&0.21$\pm$0.06&0.19$\pm$0.03\\
 8&0.005$\pm$0.000&0.86$\pm$0.24&0.39$\pm$0.06\\
 9&0.020$\pm$0.002&0.31$\pm$0.09&0.28$\pm$0.03\\
10&0.013$\pm$0.001&0.46$\pm$0.15&0.70$\pm$0.09\\
11&0.028$\pm$0.002&0.34$\pm$0.10&0.40$\pm$0.03\\
12&0.028$\pm$0.002&0.65$\pm$0.18&0.58$\pm$0.06\\
13&0.011$\pm$0.001&0.51$\pm$0.14&0.93$\pm$0.10\\
14&0.056$\pm$0.004&--&--\\
15&0.004$\pm$0.000&0.79$\pm$0.27&1.42$\pm$0.16\\
16&0.012$\pm$0.001&0.48$\pm$0.12&0.75$\pm$0.06\\
17&0.011$\pm$0.001&0.50$\pm$0.14&1.10$\pm$0.08\\
18&0.027$\pm$0.002&0.52$\pm$0.14&1.25$\pm$0.10\\
19&0.016$\pm$0.001&0.43$\pm$0.12&0.55$\pm$0.06\\
20&0.078$\pm$0.006&0.43$\pm$0.12&0.59$\pm$0.05\\
21&0.233$\pm$0.017&0.85$\pm$0.22&0.73$\pm$0.05\\
22&0.027$\pm$0.002&0.33$\pm$0.10&0.51$\pm$0.05\\
23&0.015$\pm$0.001&0.41$\pm$0.12&0.24$\pm$0.04\\
24&0.019$\pm$0.002&0.42$\pm$0.12&0.05$\pm$0.03\\
25&0.013$\pm$0.001&0.38$\pm$0.11&0.57$\pm$0.06\\
26&0.007$\pm$0.001&0.39$\pm$0.10&--\\
27&0.040$\pm$0.003&0.45$\pm$0.12&0.40$\pm$0.04\\
28&0.030$\pm$0.002&0.33$\pm$0.10&--\\
29&0.037$\pm$0.003&0.29$\pm$0.09&--\\
\hline
Seashell&0.088$\pm$0.007&0.27$\pm$0.11&1.96$\pm$0.18\\
NGC~5291&$\ge$0.264$\pm$0.021&0.42$\pm$0.18:&3.05$\pm$0.29:\\
HI concentration&0.064$\pm$0.007&--&0.77$\pm$0.17\\
\hline
\end{tabular}
\end{table*}

The star formation rate derived from the NUV summed along the entire HI structure but excluding that of NGC~5291 and the Seashell galaxies amounts to $1.37\pm0.10$ M$_\textrm{\sun}$\,yr$^{-1}$. The SFR of NGC~5291 and the Seashell is only 0.35 M$_\textrm{\sun}$\,yr$^{-1}$, or $\sim$20\% of the SFR in the entire field of view. Thus, star formation in this system is currently mostly observed in the intergalactic medium.

The total star formation rate in the regions in which we have measured fluxes in all three bands is:
\begin{itemize}
\item SFR(UV)=$\left[NUV\right]=1.31\pm0.10$ M$_\textrm{\sun}$\,yr$^{-1}$
\item SFR(H$\alpha$)=$\left[H\alpha\right]=0.71\pm0.22$ M$_\textrm{\sun}$\,yr$^{-1}$
\item SFR(8.0)=$\left[8.0\right]=0.54\pm0.02$ M$_\textrm{\sun}$\,yr$^{-1}$
\end{itemize}

The star formation rate derived from the ultraviolet is thus almost twice as large as that derived from H$\alpha$ which itself is 30\% larger than that derived using the mid-infrared.

\begin{figure}
\resizebox{\hsize}{!}{\includegraphics{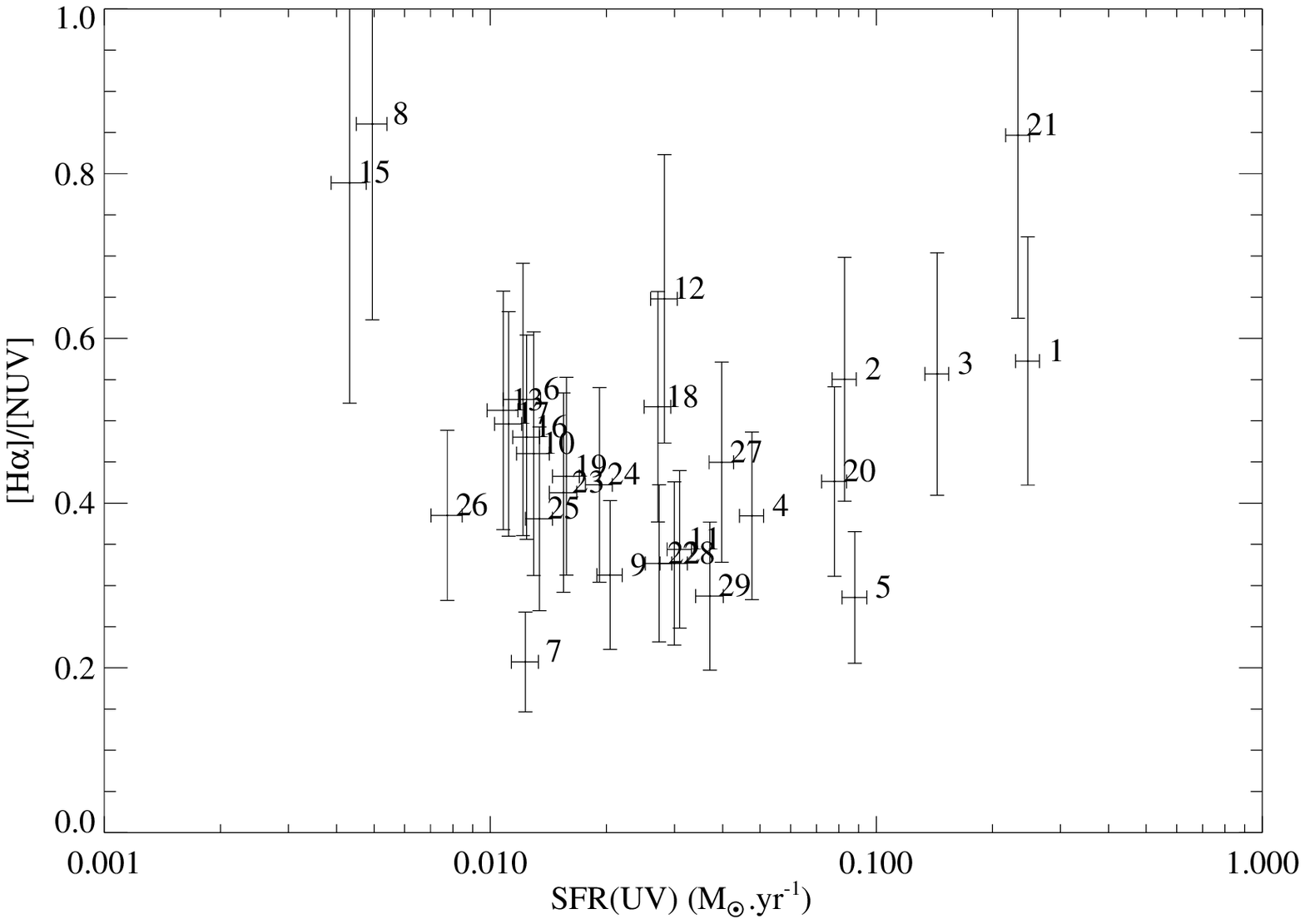}}
\caption{Ratio of $\left[H\alpha\right]/\left[NUV\right]$ versus SFR(UV).\label{fig:sfruv_vs_sfrhaovsfruv}}
\end{figure}

\begin{figure}
\resizebox{\hsize}{!}{\includegraphics{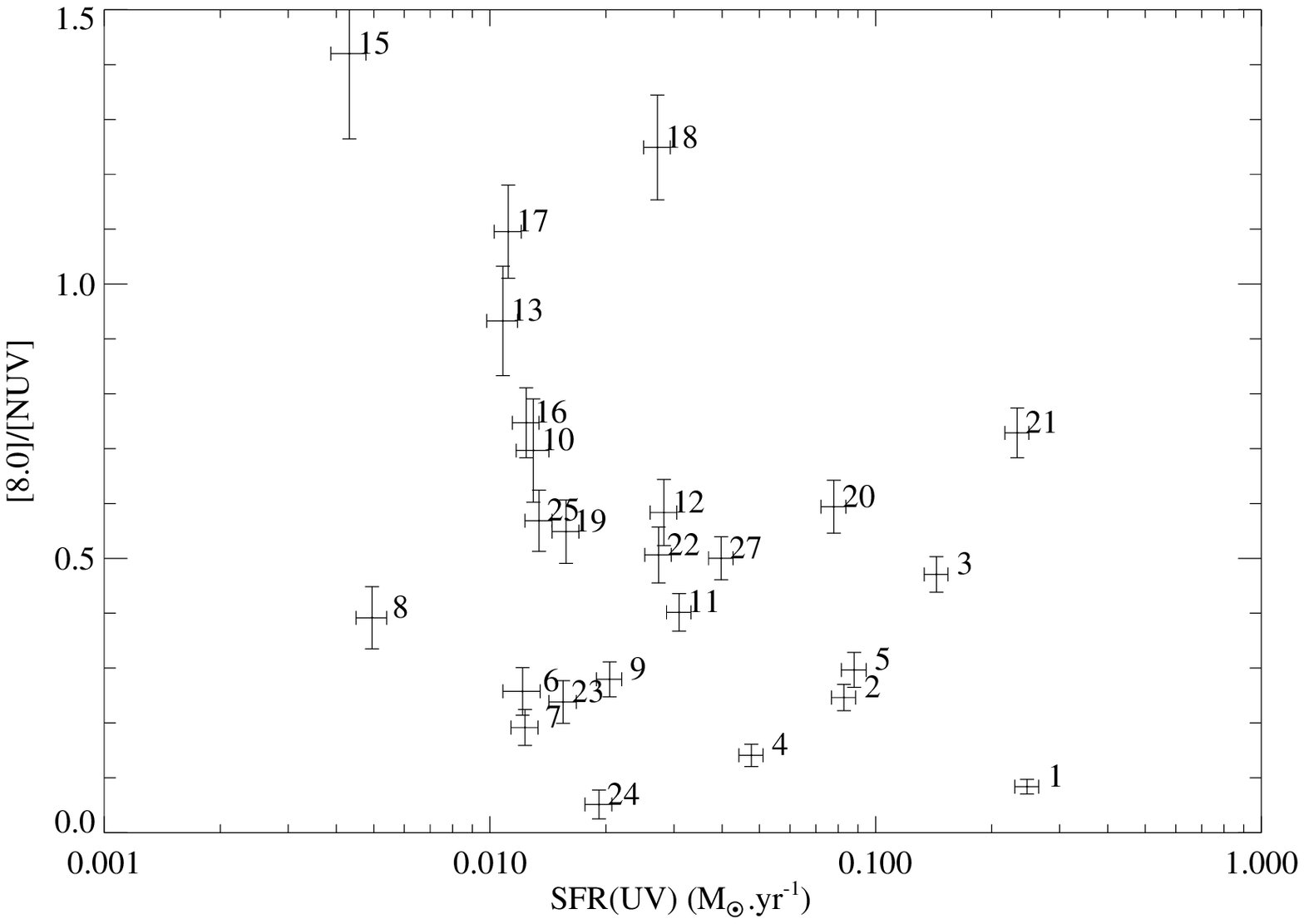}}
\caption{Ratio of $\left[8.0\right]/\left[NUV\right]$ versus SFR(UV).\label{fig:sfruv_vs_sfrirovsfruv}}
\end{figure}

To compare the different SFR estimators, we plot $\left[H\alpha\right]/\left[NUV\right]$ (Figure \ref{fig:sfruv_vs_sfrhaovsfruv}) and $\left[8.0\right]/\left[NUV\right]$ (Figure \ref{fig:sfruv_vs_sfrirovsfruv}) versus SFR(UV). On average, we get $\left[H\alpha\right]/\left[NUV\right]$=$0.47\pm0.16$ and $\left[8.0\right]/\left[NUV\right]=0.53\pm0.35$.

Figures \ref{fig:sfruv_vs_sfrhaovsfruv} and \ref{fig:sfruv_vs_sfrirovsfruv} indicate strikingly large variations of $\left[H\alpha\right]/\left[NUV\right]$ and $\left[8.0\right]/\left[NUV\right]$ from one region to an other, even among the most luminous ones. For instance, there is a factor $\sim$2 increase in $\left[H\alpha\right]/\left[NUV\right]$ when comparing region 20 with 21, or a factor $\sim$6 decrease in $\left[8.0\right]/\left[NUV\right]$ when going from region 1 to 3. Variations can also be important within a unique region, see for instance region 1 in Figure \ref{fig:vign}. As discussed in Section \ref{sec:discussion}, local variations of the star formation history, dust extinction and metallicity can in principle explain this scatter. This will be investigated further in the discussion.

\subsection{Intergalactic versus galactic star formation}

The intergalactic star-forming regions we have studied are objects intermediate between isolated star-forming dwarf galaxies and individual HII regions in spiral disks. Indeed, the most luminous of them have simultaneously the structural properties and optical colours of typical primordial blue compact dwarf galaxies and the interstellar medium content, in particular the metallicity, of their parent disks. It therefore makes sense to compare the individual SF regions around NGC~5291 with both samples of individual galaxies, studied globally, and samples of individual HII regions within nearby galaxies (M51, M81 and Arp~82). Fortunately, several {\em Spitzer} based papers have recently provided the data required for these comparisons.

\begin{figure}
\resizebox{\hsize}{!}{\includegraphics{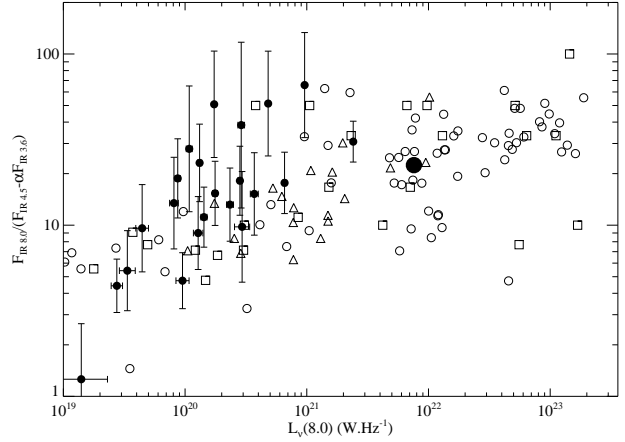}}
\caption{Ratio of the mid-infrared fluxes $\mathrm{F_{IR8.0}/(F_{IR4.5}-\alpha F_{IR3.6}})$ versus the monochromatic luminosity at 8.0~$\mu$m of SINGS galaxies \citep[][ open circles]{dale2007a}, of the dwarf galaxies in the Rosenberg et al. sample (triangles), of Engelbracht et al. sample (squares) and the HII regions in NGC~5291 (filled circles with error bars). The large filled black circle represents the sum over all NGC~5291 regions. \label{fig:sings-rosenberg-engelbracht}}
\end{figure}

We first compare the properties of the intergalactic star-forming regions in NGC~5291 with those measured globally for star-forming galaxies. For this comparison, we used three samples: the one studied by \cite{rosenberg2006a}, consisting of 19 galaxies with a B-band absolute magnitude $M_B>-18.0$; the sample of galaxies, mostly spirals, in the {\em Spitzer} SINGS legacy survey \citep{dale2007a}; and the \cite{engelbracht2005a} sample consisting of a heterogeneous collection of star forming galaxies without strong active nuclei including metal deficient blue compact dwarf galaxies. In Figure \ref{fig:sings-rosenberg-engelbracht}, we plot the 8.0~$\mu$m fluxes, normalised to the 4.5 $\mu$m fluxes (a tracer of the hot dust emission) as a function of the monochromatic luminosity at 8.0~$\mu$m. The 4.5 $\mu$m band was corrected for contamination by the photospheric stellar emission determined from the 3.6 $\mu$m band (which we assumed not to be significantly polluted by hot dust): $\mathrm{F_{IR4.5}-\alpha F_{IR3.6}}$, where $\alpha$ is a scaling coefficient. We adopted $\alpha=0.57$ as used in \cite{engelbracht2005a}. This is derived from Starburst99 models \citep{leitherer1999a} which give typical values between 0.53 and 0.61; Engelbracht assigned $\alpha$ an uncertainty of 7\%.

Most of the individual regions of NGC~5291 exhibit roughly the same range of monochromatic luminosities at 8.0~$\mu$m as the Rosenberg et al. sample of dwarf galaxies except for the brightest ones (see Figure \ref{fig:sings-rosenberg-engelbracht}). Based on their SFRs only, each of the 29 individual regions could by itself be considered a dwarf galaxy. Besides, they have on average a $\mathrm{F_{IR8.0}/(F_{IR4.5}-\alpha F_{IR3.6}})$ ratio slightly, but not significantly, higher than the dwarf galaxy sample (see Table \ref{tab:fluxratiosamples}). The ratio for the SINGS is slightly larger but very similar. The Engelbracht et al. sample, in spite of its inhomogeneity, shows a similar mean value as for the NGC~5291 regions.

\begin{figure}
\resizebox{\hsize}{!}{\includegraphics{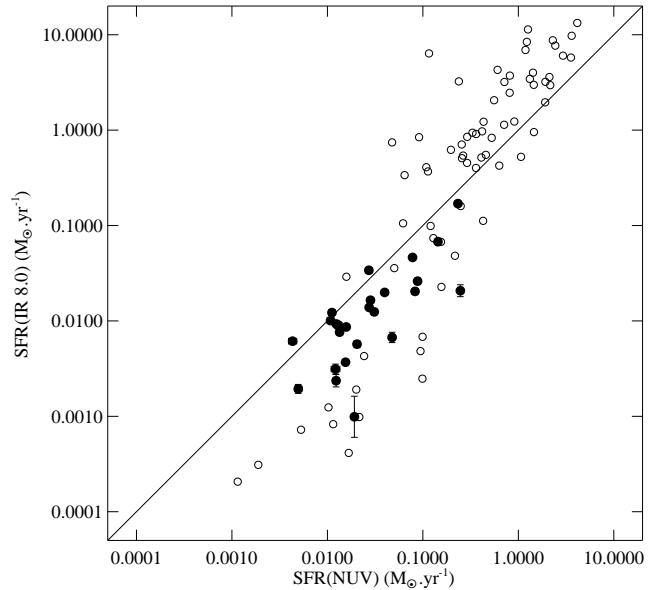}}
\caption{Star formation rate estimated from the IRAC 8.0~$\mu$m band as a function of SFR(UV) for the NGC~5291 intergalactic SF regions (filled circles) and for the galaxies of the SINGS sample \citep[][ open circles]{dale2007a}.\label{fig:comp-nuv-ir8-sings}}
\end{figure}

In figure \ref{fig:comp-nuv-ir8-sings} we compare $\left[NUV\right]$ to $\left[8.0\right]$ for the galaxies of the SINGS sample and the HII regions detected in the NGC~5291 system.

We notice that the least luminous galaxies from the SINGS sample present an important ultraviolet excess. We think it is due to the PAH defiency seen in the low metallicity galaxies, rather than an ultraviolet excess per se. On the other hand, the most luminous SINGS galaxies present a strong infrared excess, due to the higher metallicity and the large population of non-ionizing stars which can excite the PAH in addition to massive young stars.

The relative scatter (see Table \ref{tab:fluxratiosamples}) for all these samples are quite similar. This is surprising as the Rosenberg et al., the Englebracht et al. and the SINGS samples are composed of heterogeneous dwarf galaxies whereas the NGC~5291 HII regions lie in the same HI structure.

\begin{figure}
\resizebox{\hsize}{!}{\includegraphics{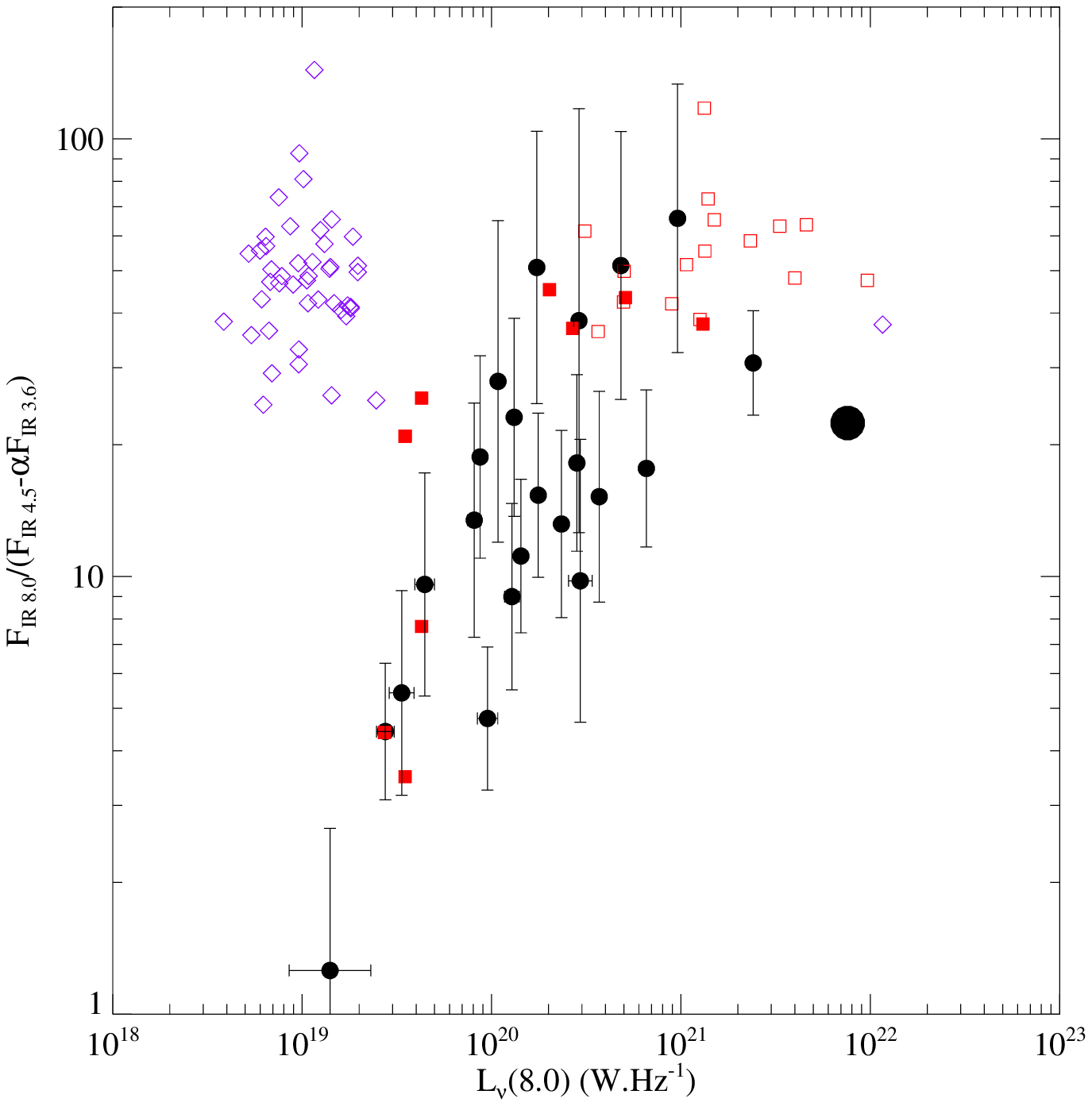}}
\caption{Ratio of the mid-infrared fluxes $\mathrm{F_{IR8.0}/(F_{IR4.5}-\alpha F_{IR3.6}})$ versus the monochromatic luminosity at 8.0~$\mu$m of M81 \citep[][ purple diamonds]{perez2006a}, Arp~82 \citep[][ red squares; filled squares represent star forming regions in tidal features]{hancock2007a} and NGC~5291 (filled circles with error bars) HII regions. The large filled circled represents the integration of all NGC~5291 regions.\label{fig:m81}}
\end{figure}

In Figure \ref{fig:m81} we perform a similar comparison as in Figure \ref{fig:sings-rosenberg-engelbracht} but with HII regions in the nearby spiral M81\footnote{The mid-infrared flux densities were provided by P\'erez-Gonz\'alez (priv.\ comm.).} \citep[for more details see][]{perez2006a} and the interacting system Arp~82 \citep{hancock2007a}.

The $\mathrm{F_{IR8.0}/(F_{IR4.5}-\alpha F_{IR3.6}})$ ratio of the regions in M81 is roughly similar to that of the most luminous HII regions of NGC~5291 even though their 8.0~$\mu$m monochromatic luminosity is lower by one to two orders of magnitude. Beside, for the less luminous regions, the ratio is higher in M81 which could indicate a deficiency in PAH emission in the HII regions of NGC~5291. The star forming regions in Arp~82, especially along the tidal arms of the interacting system, follow more closely what can be seen in NGC~5291 both in the 8.0~$\mu$m flux and the $\mathrm{F_{IR8.0}/(F_{IR4.5}-\alpha F_{IR3.6}})$. Therefore both systems have on average the same PAH properties compared to the dust continuum.

\begin{table*}
\label{tab:ratio}
\caption{Mean flux and SFR ratios for the different samples. Extremely discrepant points (if any) have been discarded before performing the calculation.}
\label{tab:fluxratiosamples}
\centering
\begin{tabular}{c c c c}
\hline\hline
Sample&$\left<\mathrm{F_{IR8.0}/(F_{IR4.5}-\alpha F_{IR3.6}})\right>$&$\left<\left[8.0\right]/\left[NUV\right]\right>$&$\left<\left[H\alpha\right]/\left[NUV\right]\right>$\\\hline
NGC~5291&$21\pm17$&$0.5\pm0.4$&$0.5\pm0.2$\\
Rosenberg&$17\pm12$&--&--\\
SINGS&$25\pm16$&$3.3\pm7.0$&--\\
Engelbracht&$24\pm23$&--&--\\
M51&--&$4.2\pm5.5$&$1.1\pm1.3$\\
M81&$50\pm35$&$1.2\pm0.9$&$2.8\pm2.2$\\
Arp~82&$45\pm24$&$1.8\pm1.8$&$1.0\pm0.7$\\
\hline
\end{tabular}
\end{table*}

\begin{figure}
\resizebox{\hsize}{!}{\includegraphics{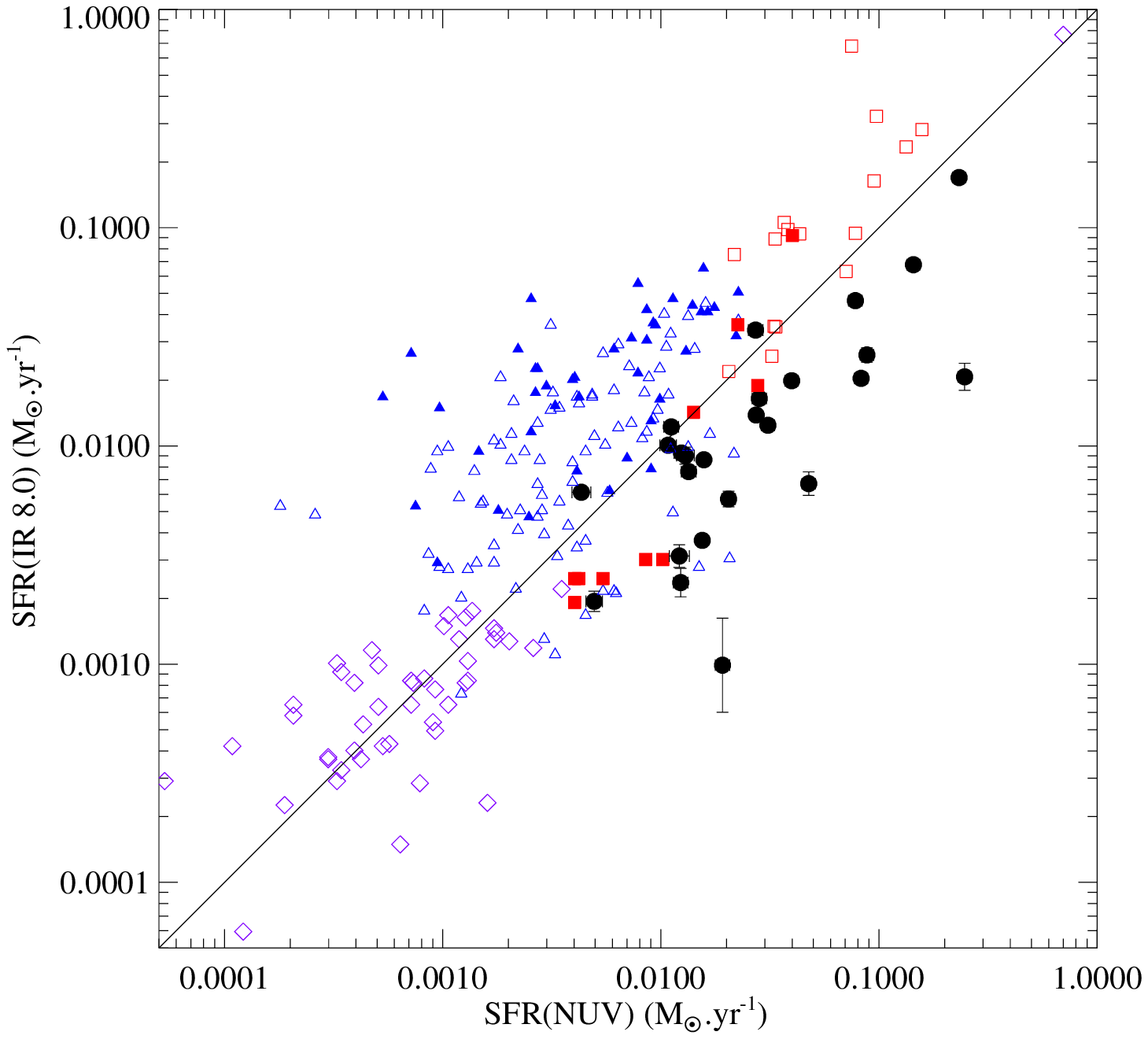}}
\caption{Star formation rate estimated from the IRAC 8.0~$\mu$m band as a function of SFR(UV) for the NGC~5291 intergalactic SF regions (filled circles), for samples of HII regions in two spiral disks: M51 \citep[][ blue triangles; filled triangles represent inner regions where Pa$\alpha$ flux is available]{calzetti2005a} and M81 \citep[][ purple diamonds]{perez2006a}, and for the interacting system Arp~82 \citep[][ red squares; filled squares represent star forming regions in tidal features]{hancock2007a}.\label{fig:comp-nuv-ir8}}
\end{figure}

\begin{figure}
\resizebox{\hsize}{!}{\includegraphics{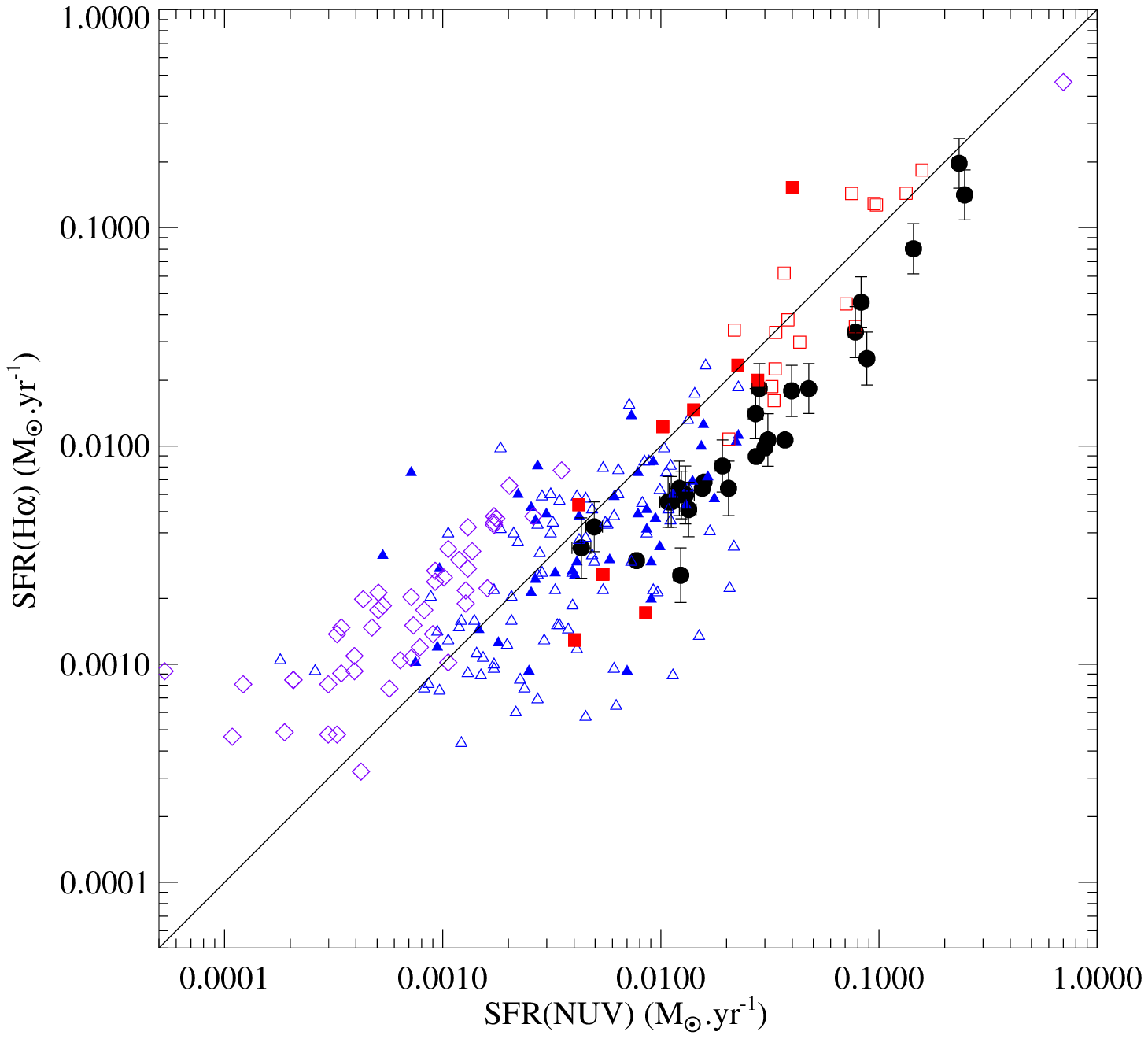}}
\caption{Star formation rate estimated from the $\left[H\alpha\right]$ luminosity as a function of SFR(UV) for the NGC~5291 intergalactic SF regions (filled circles), for samples of galactic HII regions in two spiral disks: M51 \citep{calzetti2005a} (blue triangles; filled triangles represent HII regions near centre of M51 where Pa$\alpha$ flux is available) and M81 \citep[][ purple diamonds]{perez2006a}, and for the interacting system Arp~82 \citep[][ red squares; filled squares represent star forming regions in tidal features]{hancock2007a}.\label{fig:comp-nuv-ha}}
\end{figure}

In Figures \ref{fig:comp-nuv-ir8} and \ref{fig:comp-nuv-ha}, respectively, we present a comparison of the $\left[NUV\right]$ to $\left[8.0\right]$ and $\left[NUV\right]$ to $\left[H\alpha\right]$ SFRs for individual HII regions in the HI intergalactic structure of NGC~5291, for the HII regions in the disk of M51 \citep{calzetti2005a}, M81 \citep{perez2006a} and Arp~82 \citep{hancock2007a}.

We noticed in section \ref{sec:SFRIGM} that the SFRs of the NGC~5291 HII regions estimated from the UV fluxes were on average higher by a factor of 2 than those obtained from the H$\alpha$ and 8.0~$\mu$m bands\footnote{Using the standard calibration of \cite{kennicutt1998a} and the assumptions made therein.}. Figures \ref{fig:comp-nuv-ir8} and \ref{fig:comp-nuv-ha} show that this systematic UV-excess is not observed for the individual HII regions of M51 and M81; the regions in M51 present an IR excess, whereas those of M81 seem to be too H$\alpha$ rich. No systematic UV excess can be detected in the Arp~82 interacting system.

More quantitatively, when normalised to the 8.0~$\mu$m band, the NUV flux in NGC~5291 is in excess by a factor of 8, 2 and 1.6 compared to M51, M81 and Arp~82 respectively. When normalised to H$\alpha$, the corresponding excess factors are 2, 6 and 2. Finally, one should note that NGC~5291 presents a scatter in the H$\alpha$ vs.\ UV relation which is smaller than for the comparison galaxies, while the MIR vs.\ UV is roughly comparable.

In summary, while globally the 8.0~$\mu$m emission in the intergalactic star forming regions of NGC~5291 are not significantly different than those of star-forming dwarf galaxies, they seem to present with respect to individual HII regions in spirals, and with the H$\alpha$ and MIR SF indicators, a clear UV excess.

\section{Discussion\label{sec:discussion}}

In the previous section, we presented the properties of the star forming regions along the huge HI structure around NGC~5291, a study based on three SF indicators: UV, H$\alpha$ and the mid-infrared. We found that the three of them were correlated, as for more classical galactic star-forming regions, but showed with the latter some differences, in particular a UV excess. We discuss here the origins of these differences and of the scatter in the correlations. The goal is to determine which SF indicator -- or combination of them -- may be used to estimate the total SFR in this special environment, bearing in mind that the previous calibrations were derived for galactic environments. Once a ``total'' SFR is determined, we briefly discuss its impact on the IGM.

\subsection{The effect of metallicity and PAH strength on the MIR star formation indicator}

In most galactic star-forming regions, the dust emission around 8.0~$\mu$m is largely dominated by PAH features. The latter are however known to be absent in metal-poor objects such as the Blue Compact Dwarf Galaxies \citep{galliano2005a,engelbracht2005a,madden2006a,wu2006a}. Thus for these systems, the IRAC based mid-IR--SFR calibration, as the one used in this study, could be fairly inaccurate. The intergalactic HII regions around NGC~5291, although they share many properties with those of BCDGs, lie however within an HI structure which had been previously pre-enriched. \cite{duc1998a} measured an oxygen abundance of $12+\log\left(\textrm{O/H}\right)=8.4$ to $8.6$, with an average of $8.49^{+0.07}_{-0.10}$. Although this is somewhat below the solar value, this is consistent with a PAH detection. Indeed, the IRS spectra obtained by \cite{higdon2006a} indicate that the IRAC 8.0~$\mu$m emission of the two most luminous star forming regions is mainly due to PAHs.

\begin{figure}
\resizebox{\hsize}{!}{\includegraphics{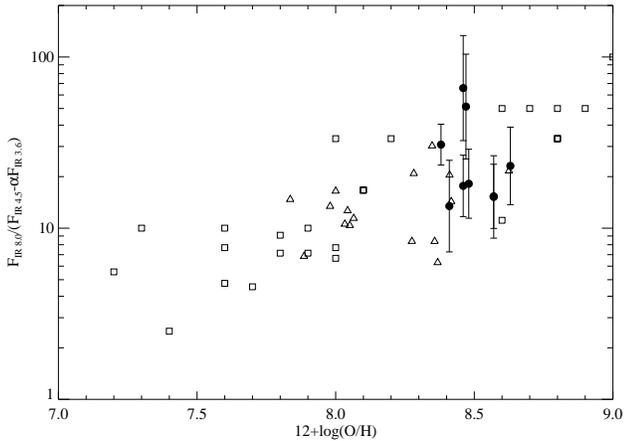}}
\caption{Ratio of the mid-infrared fluxes $\mathrm{F_{IR8.0}/(F_{IR4.5}-\alpha F_{IR3.6}})$ (with $\alpha=0.57$) versus the metallicity of dwarf galaxies from the Rosenberg et al. sample (triangles), Engelbracht galaxies (open squares) and HII regions in NGC~5291 (filled circles).\label{fig:rosenberg-metal-36-80}}
\end{figure}

In Figure \ref{fig:rosenberg-metal-36-80}, we plot the PAH emission normalised to the dust continuum as a function of metallicity for the NGC~5291 star forming regions, the dwarf galaxies from the Rosenberg sample and the galaxies from the Engelbracht sample. The luminosity at 8.0~$\mu$m appears to be consistent with that of dwarf galaxies with the same metallicity. \cite{wu2006a} directly measured a correlation between the equivalent widths of the PAH features and metallicity. We checked this for region 21. We found that the EWs measured in the IRS spectrum published in \cite{higdon2006a}\footnote{EW(6.2 $\mu$m)=0.58 $\mu$m, EW(7.7 $\mu$m)=0.52 $\mu$m, EW(8.6 $\mu$m)=0.53 $\mu$m, EW(11.2 $\mu$m)=0.65 $\mu$m.}, using a method similar as in the \cite{wu2006a} paper, were consistent with the Wu et al. correlation.

Thus, on average, the relative strength of the 8.0~$\mu$m emission in NGC~5291 appears to be normal, i.e. consistent with that expected for star-forming regions with the same metallicity. Therefore the calibration derived between the IRAC mid-infrared emission and the SFR so far only validated for spiral disks should also apply to the collisional debris.

The large scatter in the $\mathrm{F_{IR8.0}/(F_{IR4.5}-\alpha F_{IR3.6}})$ flux ratio along the system (see Figures \ref{fig:sings-rosenberg-engelbracht} and \ref{fig:m81}) is however surprising. In any case, it cannot be simply explained by local variations of the metallicity as the latter was measured by \cite{duc1998a} to be quite uniform all along the HI structure. For instance, very bright regions, such as 1 and 3, show discrepant $\mathrm{F_{IR8.0}/(F_{IR4.5}-\alpha F_{IR3.6}})$ flux ratios although they lie very close on the sky and thus should have very similar metallicities. Unfortunately no optical spectrum of region 1 is available to check this. The IRS spectra of the two most luminous regions (regions 3 and 21 in this paper) also show variations in their PAH strengths and dust continuum shortwards of 5 $\mu$m and above 9 $\mu$m, which are not yet understood. Clearly, the metallicity is not the only parameter playing a role in the 8.0~$\mu$m emission, as also shown in several studies of galactic star formation.

\subsection{The effect of dust extinction on the UV and H$\alpha$ star formation indicator}
We have seen earlier that dust is undoubtedly present all along the HI ring. Could local variations of the dust extinction and geometrical factors account for the large scatter in the individual SF properties? To examine that, we have corrected both $\left[NUV\right]$ and $\left[H\alpha\right]$ for extinction, the value of which was determined locally towards each condensation. The extinction is actually composed of two components: one due to the dust clouds surrounding newborn stars, which is due to the molecular gas and very dependent on geometry and may vary dramatically from one region to the other; a second one is due to the dust along the line of sight, which can be approximated by a dust screen. Assuming the ratio of dust to HI is constant, we can derive the extinction variation from the HI maps.

To quantify this HI based extinction, we used previous extinction estimates determined by \cite{duc1998a} from the measurement of the Balmer decrement in several associated HII regions. Correlating the peak HI column densities and the optical extinction in the V band with 7 data points (after having excluded 2 which showed large discrepancies), we obtained a calibration of the HI-based extinction. We estimate the uncertainty to be about 0.4 dex with this method.

To convert the extinction from the V band to other bands, we used the average Large Magellanic Cloud $A\left(\lambda\right)/A_V$ of \cite{gordon2003a}. We find $A_{NUV}/A_V=2.80$ and $A_{H\alpha}/A_V=0.89$.

\begin{figure}
\resizebox{\hsize}{!}{\includegraphics{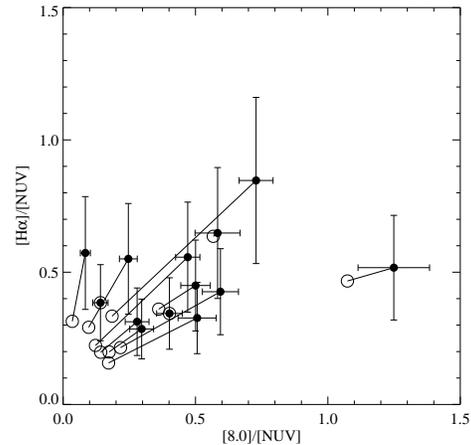}}
\caption{$\left[H\alpha\right]/\left[NUV\right]$ as a function of $\left[8.0\right]/\left[NUV\right]$ for HII knots with an uncorrected ultraviolet star formation rate greater than 0.02 M$_\textrm{\sun}$\,yr$^{-1}$. Filled circles are uncorrected for extinction while empty ones are.\label{fig:sfrir80ovsfruv_vs_sfrhaovsfruv_corr_ext}}
\end{figure}

We have corrected the NUV and H$\alpha$ fluxes for dust extinction based on the calibrated HI column densities. Results of this correction are shown in Figure \ref{fig:sfrir80ovsfruv_vs_sfrhaovsfruv_corr_ext} for objects with uncorrected ultraviolet star formation rate greater than 0.02 M$_\textrm{\sun}$\,yr$^{-1}$, which have the most secure determination of the SFR.

The relative scatter of $\left[H\alpha\right]/\left[NUV\right]$ and $\left[8.0\right]/\left[NUV\right]$ is not reduced by the extinction correction. Clearly, variations of the dust extinction alone cannot explain all the scatter in the SF properties found along the HI structure, unless very local variations dominate.

\subsection{The effect of starburst age on the UV and $H\alpha$ star formation indicators\label{sec:ageeffects}}

The so-called ``UV--excess'' seems to be a determining characteristic of the intergalactic star forming regions around NGC~5291. This was also observed in a few other intergalactic SF regions \citep{werk2006a}. We discuss here whether this excess, as well as the large scatter of the $\left[H\alpha\right]/\left[NUV\right]$ after the correction for dust extinction, could result from some starburst age effects.

\cite{duc1998a} measured H$\beta$ equivalent widths as large as 140 \AA\ of several HII regions around NGC~5291, which is indicative of young starbursts \citep[see for instance][]{terlevich2004a}. The detection of Wolf-Rayet features in the most luminous one shows that the current star formation episode there is less than $5\times10^6$ years old. More recently, \cite{higdon2006a} determined ages of $\sim5\times10^6$ years from \mbox{[Ne II]/[Ne III]} emission line ratios for the objects for which mid-infrared IRS spectra were available. Are these indications for starbursts being young consistent with the UV excess observed in all intergalactic HII regions around NGC~5291? At first sight, one may think that a UV excess, or conversely, an H$\alpha$ deficit would rather suggest older, declining bursts since the UV time scale is a factor of ten larger than the H$\alpha$ one (see section \ref{sec:SFRIGM}).

\begin{figure}
\resizebox{\hsize}{!}{\includegraphics{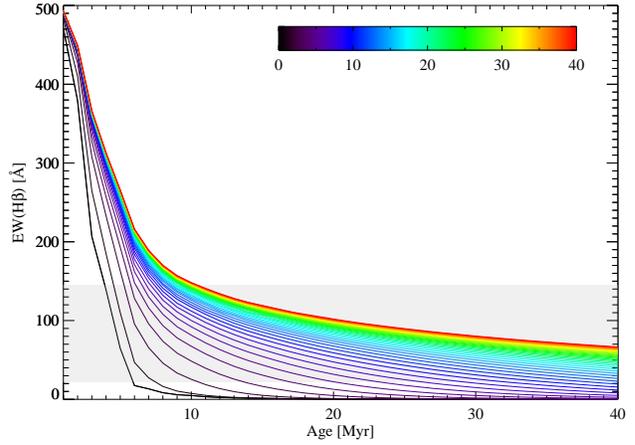}}
\caption{Plot of the variation of the equivalent width of H$\beta$ in \AA, as a function of ages for different types of starbursts. The black line shows an instantaneous starburst whereas the red one corresponds to an exponentially decreasing starburst with a timescale $\tau=40\times10^6$ years. Lines for all intermediate $\tau$ with intervals of $1\times10^6$ years are also shown. The light grey zone indicates the range of observed values of the H$\beta$ equivalent width for the few regions where this was measured \citep{duc1998a}.\label{fig:t_vs_whb_pegase}}
\end{figure}

\begin{figure}
\resizebox{\hsize}{!}{\includegraphics{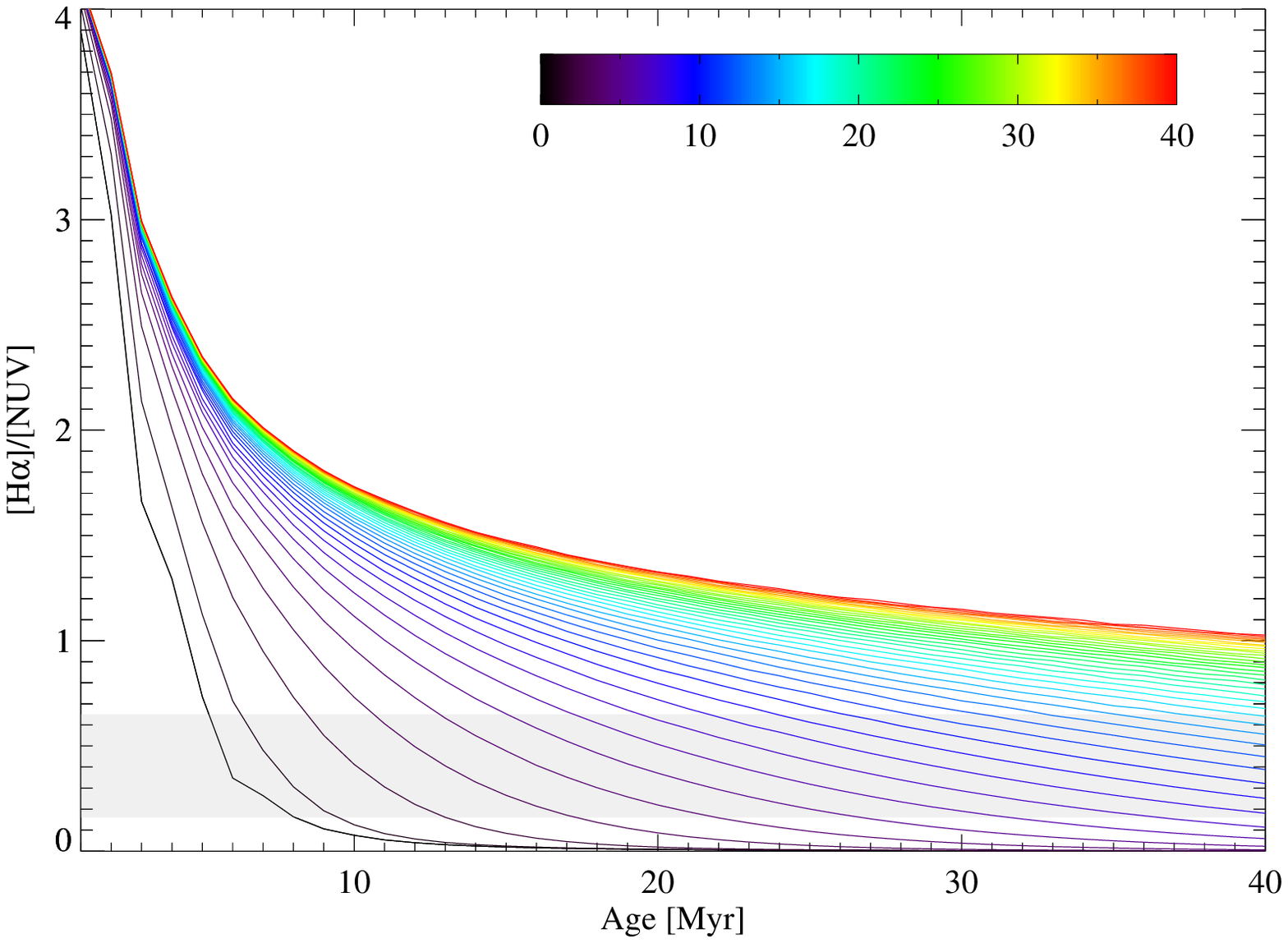}}
\caption{Plot of the variation of the $\left[H\alpha\right]/\left[NUV\right]$ ratio as a function of age for different types of starbursts. The black line shows an instantaneous starburst whereas the red one corresponds to an exponentially decreasing starburst with a timescale $\tau=40\times10^6$ years. Lines for all intermediate $\tau$ with intervals of $1\times10^6$ years are also shown. The light grey zone indicates the range of observed values of $\left[H\alpha\right]/\left[NUV\right]$ after correction for extinction. \label{fig:t_vs_sfrhaovsfruv_pegase}}
\end{figure}

To pursue the analysis, we have modelled the HII regions using the spectral evolution code {\sc P\'egase II} \citep{fioc1997a,fioc1999a}. We have used a set of parameters differing by the star formation history (SFH), going from an instantaneous starburst to exponentially decreasing ones with a timescale between 1 and 40 million years. We used a Salpeter IMF from 0.1 to 100 M$_\textrm{\sun}$ and a constant metallicity Z=0.008 (65\% solar metallicity, corresponding to $12+\log{(O/H)}=8.47$), in agreement with the metallicity determined by \cite{duc1998a}. Extinction from the gas has not been taken into account in the simulations in order to isolate age effects only. For comparison we therefore need to use the extinction corrected ratios. We have in particular studied how the H$\beta$ equivalent width and the $\left[H\alpha\right]/\left[NUV\right]$ ratio evolve with SFH (see Figures \ref{fig:t_vs_whb_pegase} and \ref{fig:t_vs_sfrhaovsfruv_pegase}). The first plot suggests that the ongoing star formation may not have lasted for more than $10\times10^6$ years for the regions presently showing the largest H$\beta$ equivalent width, whatever the timescale assumed for the starburst. Other than that, very young, quasi-instantaneous starbursts ignited $3-5\times10^6$ years ago may well cohabit with older and more extended star-formation episodes \citep[][]{neff2005a}, i.e. in regions showing smaller H$\beta$ equivalent width. The second plot shows that the observed range for $\left[H\alpha\right]/\left[NUV\right]$ may be obtained for various SFH, but that in any case, all bursts are well advanced. A constant or just slowly declining SFR is inconsistent with the data. However, our dataset may accommodate quasi-instantaneous bursts which occurred $4-10\times10^6$ years ago up to bursts with slightly longer time scales of $10\times10^6$ years which started about $40\times10^6$ years ago. Combining the two criteria, we conclude that the regions that are currently the most active ones (i.e. region 1, 3 and 21) are all very young but should have had their SF peak already $5\times10^6$ years ago while the more quiescent ones may be a few $10\times10^6$ years older, still sustaining star formation but at a lower rate than before.

One should note that quasi-instantaneous young starbursts for all HII regions around NGC~5291 are also fully compatible with the data, i.e. the H$\beta$ equivalent width, the UV excess and the somehow large region-to-region variations of the $\left[H\alpha\right]/\left[NUV\right]$ to UV ratio. The latter may be understood noting how this ratio changes quickly after the beginning of the starburst: a small age difference can produce a considerable difference in $\left[H\alpha\right]/\left[NUV\right]$ (see Figure \ref{fig:t_vs_sfrhaovsfruv_pegase}). Probing the presence, or actually the absence, of an intermediate age stellar population of a few $10\times10^6$ years would be required to validate the scenario of a rapidly decreasing starburst. This requires a model of the full Spectral Energy Distribution of each region, from the UV to IR, which is beyond the scope of this study. This task is somehow easier in the case of the NGC~5291 intergalactic star-forming regions. Indeed, they should not contain a stellar component of a few $100\times10^6$ years or older coming from the parent galaxies.

As all the HII regions seem to have roughly the same age it is legitimate to wonder how star-formation may have started quasi simultaneously along a structure as extended as 180 kpc, and how likely we are to observe it so close to its onset. One should however remember that the HI ring which hosts all these regions was probably formed during a single violent event, likely the collision of a previously existing extended HI disk with a companion galaxy \citep{duc1998a}. Given its ring-like shape and overall velocity field, the HI structure was most probably not formed by tidal forces but rather by an expanding density wave following a bull's eye collision. According to numerical simulations \citep{bournaud2007a}, the ring around NGC~5291 has an age between $300\times10^6$ and $400\times10^6$ years, very different from the star formation timescale. However, in the case of a collisional ring and unlike tidal tails, the star formation can be suppressed during the first few $100\times10^6$ years, i.e., until the HI ring expansion stalls after which star formation proceeds. If that scenario is indeed correct, this would bring the dynamical and the star-formation time scales together and explain the, at first sight, rather intriguing observation of quasi simultaneous star formation along the entire ring.

\subsection{The total intergalactic star formation rate in NGC~5291: impact on the IGM\label{section:sfr}}

The integrated star formation rate along the HI structure -- uncorrected for dust extinction -- derived from the GALEX NUV emission, amounts to 1.31$\pm$0.10 M$_\textrm{\sun}$\,yr$^{-1}$, 0.71$\pm$0.22 M$_\textrm{\sun}$\,yr$^{-1}$ in H$\alpha$ and 0.54$\pm$0.02 M$_\textrm{\sun}$\,yr$^{-1}$ from the 8.0~$\mu$m band emission (see section \ref{sec:SFRIGM}).

Whether the uncorrected UV, H$\alpha$ and IR SFRs should be added to get the total, dust free SFR, or whether a simple extinction correction for each tracer is enough is a matter of active debate \citep{hirashita2003a}. Because our estimates of the extinction, based on the HI column density, are rather uncertain, we decided not to apply a dust correction for each individual region. Instead we decided to add the uncorrected SFR obtained from H$\alpha$ to the SFR based on the dust emission scaled by a coefficient which takes into account the amount of ionizing radiation (actually radiation emitted by the same stars as traced through H$\alpha$) absorbed by PAH. The latter was determined empirically by fitting SFR(H$\alpha$)+k$\times$SFR(IR) to the extinction corrected SFR(H$\alpha$) using only datapoints for which measures of $A_V$ obtained from the Balmer decrement were available. The advantage of using a combination of H$\alpha$ and IR rather than UV and IR (as commonly done) is that the former two indicators probe SF with relatively similar time scales of $10\times10^6$ years (rather than $100\times10^6$ years for UV). Doing so, we obtained $k=0.27$ and a total SFR all along the HI structure of 0.83 M$_\textrm{\sun}$\,yr$^{-1}$. This value is a factor of 1.7 smaller than the value determined from the uncorrected near UV emission, confirming again the earlier mentioned UV excess.

However, we should note at this stage that the SFRs were estimated using calibrations valid for a constant star formation over $100\times10^6$ years \citep{kennicutt1998a}. As mentioned before, the star formation history in NGC~5291 may have been quite different. There are various independent indications pointing towards young, quasi-instantaneous, starbursts and the measure of the actual SFR may be different.

\begin{figure}
\resizebox{\hsize}{!}{\includegraphics{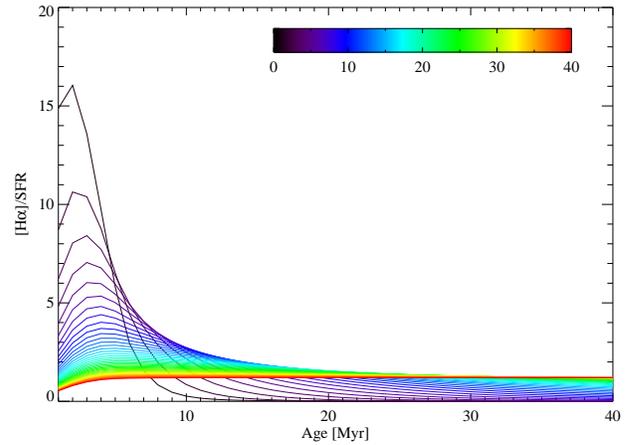}}
\caption{Plot of the variation of $\left[H\alpha\right]$ (it is to say the star formation rate that would give the Kennicutt estimator) over the actual star formation rate as a function of age for different types of starbursts. The black line shows an instantaneous starburst whereas the red one corresponds to an exponentially decreasing starburst with a timescale $\tau=40\times10^6$ years. Lines for all intermediate $\tau$ with intervals of $1\times10^6$ years are also shown.\label{fig:t_vs_sfrhaovsfrreal}}
\end{figure}

\begin{figure}
\resizebox{\hsize}{!}{\includegraphics{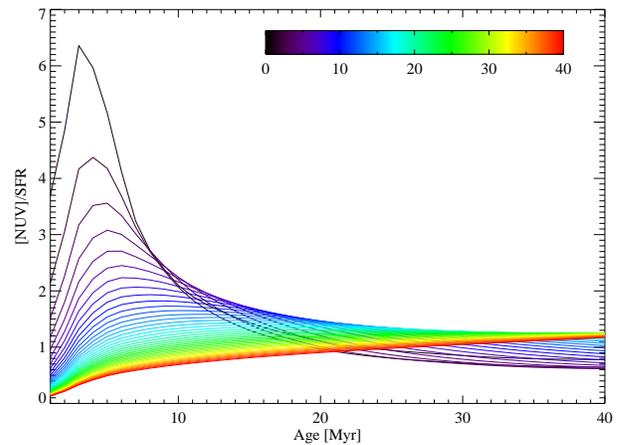}}
\caption{Plot of the variation of the $\left[NUV\right]$ (it is to say the star formation rate that would give the Kennicutt estimator) over the actual star formation rate as a function of age for different types of starbursts. The black line shows an instantaneous starburst whereas the red one corresponds to an exponentially decreasing starburst with a timescale $\tau=40\times10^6$ years. Lines for all intermediate $\tau$ with intervals of $1\times10^6$ years are also shown.\label{fig:t_vs_sfruvovsfrreal}}
\end{figure}

For young and rapidly decreasing starbursts, models (see Figures \ref{fig:t_vs_sfrhaovsfrreal} and \ref{fig:t_vs_sfruvovsfrreal}) indicate that the Kennicutt UV and H$\alpha$ calibrations overestimate by factors of 2 to 4 the actual SFR. Thus if the quasi-instantaneous starburst hypothesis is correct, the real ``intergalactic'' star formation rate around NGC~5291 should be 0.3--0.7 M$_\textrm{\sun}$\,yr$^{-1}$, still higher than for typical dwarf and irregular galaxies \citep{hunter2004a}. However, one should keep in mind that the UV-excess indicates that the star formation rate was higher a few $10^6$ years ago, with values close to that observed in quiescent spiral galaxies. 

We thus find that around NGC~5291, star-formation proceeds at a rate typical of spirals but in an environment which is much less dense, and where the effect of stellar winds on the surrounding medium should be much more dramatic. As shown, among others, by \cite{ferrara2000a}, the ability of an object to lose its heavy elements, and thus enrich the intergalactic medium, depends very much on the mass and shape of the halo it sits in. Collisional debris being dark matter deficient, they should have a reduced capability to retain their metals. How the particularly active star-forming regions in NGC~5291 manage to survive this feedback of mechanical energy still remains to be studied. Given their large gas reservoir, it is somewhat surprising that they are currently already fading, unless SF has ceased as a result of feedback from massive stars.

Clearly, a system like NGC~5291 is exceptional in the local Universe. Studying a sample of interacting systems \citep{struck2006a,smith2007a,hancock2007a} found that their external plumes and tails only account for at most 10\% of the total star formation. In NGC~5291, from the UV band we can see that more than 80\% of the stars are currently formed along the intergalactic ring. When using the 8.0~$\mu$m band, removing the old stellar content from the NGC~5291 galaxy (this population is responsible for roughly 50\% of the 8.0~$\mu$m flux), the intergalactic SF still accounts for more than 50\% of the SFR. In the distant Universe, however, such large gaseous structures may have been more common\footnote{The gigantic Ly$\alpha$ blobs found in the distant Universe \citep{nilsson2006a,jimenez2006a} actually present some similarities with NGC~5291.} and the star formation regions they hosted may have contributed significantly to the direct enrichment of the IGM.

\section{Conclusion}

We have performed a multi-wavelength study of the NGC~5291 HII regions using ground-based, GALEX, and {\em Spitzer Space Telescope} data of 29 intergalactic HII regions around NGC~5291. These regions lie along a previously identified huge HI ring-like structure of collisional origin. We have measured their fluxes in the near ultraviolet, H$\alpha$, and mid-infrared in order to compare these different star formation tracers. The morphology observed in the three bands is remarkably similar, indicating that each can serve as a star formation tracer. We have shown that the intergalactic IR emission at 8.0~$\mu$m, dominated by PAH, when normalised to the hot dust continuum, is comparable to the integrated emission of dwarf galaxies of the same metallicity and to the emission of individual HII regions in spirals. This indicates that the 8.0~$\mu$m band emission is probably an estimator of the SFR that is as reliable for these regions as it is for spirals. Whereas the 8.0~$\mu$m is ``normal'', there is a clear excess of near ultraviolet emission compared to individual HII regions in spirals. The $\left[8.0\right]/\left[NUV\right]$ and $\left[H\alpha\right]/\left[NUV\right]$ are low although there are some strong variations from one region to an other. The variations of the metallicity sensitive $\left[8.0\right]/\left[NUV\right]$ ratio cannot be explained by a metallicity effect since the oxygen abundance was found to be uniform along the entire HI structure. Moreover, correcting for the spatial variations of the dust extinction does not reduce the scatter in the $\left[H\alpha\right]/\left[NUV\right]$ and $\left[8.0\right]/\left[NUV\right]$ SFR ratios. Their variations are best explained by age effects. A model of the evolution $\left[H\alpha\right]/\left[NUV\right]$ with time favours young instantaneous starbursts: the ultraviolet excess or deficit of H$\alpha$ plus the large scatter indicating rapid variations of the $\left[H\alpha\right]/\left[NUV\right]$ ratio are difficult to explain with a constant SFR or moderately decreasing SFR. A young starburst is consistent with constraints from the H$\beta$ equivalent widths and mid-infrared NeII fluxes. So far no indication for the presence of an old stellar component has been found. A detailed study of the full SED of the star-forming regions, including dust models, would be required to confirm it. Comparing the individual starburst ages with the dynamical timescale for the formation of the HI ring in which they lie, we concluded that the apparent observation of a quasi-simultaneous onset of star formation over the 180 kpc long gaseous structure can be understood if SF commences only after expansion of the ring has stalled, after several $100 \times 10^6$\,yr.

Using the standard Kennicutt calibrations, we estimated from the UV, uncorrected for dust extinction, a total SFR around NGC~5291 of 1.3 M$_\textrm{\sun}$\,yr$^{-1}$. If the quasi-instantaneous starburst hypothesis is correct, then this current value of the SFR may have been overestimated by factors of 2 to 4. In that case, the SFR would have been much higher in the recent past. Finally it appears that about 80\% of star formation in the NGC~5291 system takes place in the intergalactic HII regions, a value much higher than in nearby interacting systems.

\begin{acknowledgements}
This research has made use of the NASA/IPAC Extragalactic Database (NED) which is operated by the Jet Propulsion Laboratory, California Institute of Technology, under contract with the National Aeronautics and Space Administration.

We thank Pablo P\'erez-Gonz\'alez, Daniel Dale and Mark Hancock for providing additional data on the comparison galaxies. We also thank our referee, Claudia Mendes de Oliveira, as well as V\'eronique Buat, Fr\'ed\'eric Bournaud and Fr\'ed\'eric Galliano for useful comments and enlightening discussions.

UL acknowledges support by the DGI grants AYA 2005-07516-C02-01 and ESP 2004-06870-C02-02, and by the Junta de Andaluc\'ia (Spain).
\end{acknowledgements}

\bibliographystyle{aa}
\bibliography{6692}
\clearpage
\end{document}